\documentclass[aps,twocolumn,showpacs,email,amssymb,nobibnotes,pre]{revtex4-1}
\usepackage{amssymb,amsmath}
\usepackage{xcolor}
\usepackage{graphicx}
\usepackage[caption=false]{subfig}
\usepackage{mathtools}
\usepackage{natbib}
\usepackage{epstopdf}
\usepackage{algorithm,algpseudocode}

\renewcommand{\l}{\left}
\renewcommand{\r}{\right}
\newcommand{\subeq}[2]{\begin{subequations}\label{#2}\begin{align}#1\end{align}\end{subequations}}
\newcommand{\eq}[2]{\begin{equation}#1\label{#2}\end{equation}}
\newcommand{\al}[2]{\begin{align}#1\label{#2}\end{align}}

\newcommand{\dd}{\text{\,d}}

\definecolor{Purpleee}{RGB}{91,40,224}

\begin{document}
\title{Accelerated Bayesian inference of gene expression models from \\ snapshots of single-cell transcripts}
\author{Yen Ting Lin}
\affiliation{Theoretical Division and Center for Nonlinear Studies, Los Alamos National Laboratory, Los Alamos, NM 87545, USA}
\author{Nicolas E. Buchler} 
\affiliation{Department of Molecular Biomedical Sciences, North Carolina State University, Raleigh, NC, 27606, USA}

\begin{abstract}
Understanding how stochastic gene expression is regulated in biological systems using snapshots of single-cell transcripts requires state-of-the-art methods of computational analysis and statistical inference.  A Bayesian approach to statistical inference is the most complete method for model selection and uncertainty quantification of kinetic parameters from single-cell data.  This approach is impractical because current numerical algorithms are too slow to handle typical models of gene expression. To solve this problem, we first show that time-dependent mRNA distributions of discrete-state models of gene expression are dynamic Poisson mixtures, whose mixing kernels are characterized by a piece-wise deterministic Markov process. We combined this analytical result with a kinetic Monte Carlo algorithm to create a hybrid numerical method that accelerates the calculation of time-dependent mRNA distributions by 1000-fold compared to current methods.  We then integrated the hybrid algorithm into an existing Monte Carlo sampler to estimate the Bayesian posterior distribution of many different, competing models in a reasonable amount of time. We validated our method of accelerated Bayesian inference on several synthetic data sets. Our results show that kinetic parameters can be reasonably constrained for modestly sampled data sets, if the model is known \textit{a priori}. If the model is unknown,the Bayesian evidence can be used to rigorously quantify the likelihood of a model relative to other models from the data. We demonstrate that Bayesian evidence selects the true model and outperforms approximate metrics, e.g., Bayesian Information Criterion (BIC) or Akaike Information Criterion (AIC), often used for model selection. 
\end{abstract}


\maketitle

\section{Introduction}
Gene expression is a biochemical process driven by the chance collisions of molecules, which can result in strong stochastic signatures and cell-to-cell variability in gene dynamics. Advances in single-cell and single-molecule technologies have provided unprecedented resolution on the stochastic dynamics of gene expression \cite{pichon2018}.  Dynamic assays measure gene expression in living cells either directly via transcript tagging \cite{golding2005,larson2011,corrigan2016,fukaya2016} or indirectly via fluorescent or luminescent proteins \cite{ozbudak2002,elowitz2002,suter2011,nicolas2018}.  Static assays measure transcript levels in fixed cells either using a cocktail of fluorescently-labelled DNA oligos that bind specific transcripts \cite{raj2008,zenklusen2008} or via single-cell RNA sequencing \cite{tang2009,hwang2018}. Static assays are popular because they do not require genetic modifications and are easily multiplexed. The disadvantage is that static assays only provide population snapshots of transcripts levels and cannot follow the dynamics of transcription in a single cell through time. 

To this end, static assays have relied upon mathematical models to {\it infer} dynamic properties of gene expression in single cells from the measured snapshot of transcript levels; see \cite{munsky2015} for a review. Inference requires (1) appropriate models of stochastic gene expression, (2) numerical methods to calculate the time-dependent mRNA distribution in a population of cells given {\it any} underlying model and associated parameters, and (3) calculating the likelihood that measured data were sampled from the calculated distribution. We recently developed a Bayesian approach (BayFISH) that uses this likelihood to infer best-fitting parameters from single cell data and quantify their uncertainty using the posterior distribution \cite{Gomez-Schiavon2017}.  Although Bayesian inference is the most complete and rigorous approach, it requires significantly more computation than other approximate methods, e.g. maximum likelihood.  

We have developed a hybrid numerical method that accelerates the calculation of time-dependent mRNA distributions by 1000-fold compared to standard methods. We integrated this method into BayFISH and, for the first time, one can estimate the Bayesian posterior distribution of many competing models in a reasonable amount of time. The Bayesian evidence rigorously quantifies the likelihood of a model relative to other models given the data, and we show that HME selects the true model and outperforms approximate metrics, e.g., BIC or AIC, typically used for model selection. Our accelerated Bayesian inference represents a significant advance over existing methods used for inferring gene expression models from snapshots of single cell transcripts. 

\section{Results}

Our inference method uses data from single-molecule RNA Fluorescence In Situ Hybridization (smFISH), but could include single cell data from other static assays. The smFISH technique labels transcripts with fluorescent DNA oligos and measures both the number of mature mRNAs ($m$) and the number of gene loci with high-activity transcription sites ($TS$); see Figure 1a. A typical smFISH data set is a histogram $h = h(\Vec{\omega})$ where $\Vec{\omega}\in \Omega$ is the set of all possible states $(m,TS)$ that can be measured in a cell.

\subsection{Connecting models of gene expression to single cell data}

A broad spectrum of measured gene expression profiles in bacteria and eukaryotes is well-explained by discrete state gene expression models \cite{rajcell2008,nicolas2017}, summarized by the following reactions:
\begin{eqnarray}
{\rm Gene \; state \;}i & \xrightarrow{\text{$\kappa_{ij}$}} & {\rm Gene \; state \;}j, \nonumber \\
{\rm Gene \; state \;}i & \xrightarrow{\text{$\beta_i$}} & {\rm Gene \; State \;}i + {\rm mRNA} \nonumber, \\
{\rm mRNA} & \xrightarrow{\text{$\delta$}} & \oslash.  
\end{eqnarray}
In this article, we adopt a two-allele, 3-state model (Figure 1b) as a case study for modelling stochastic gene expression in eukaryotes and for testing our method of accelerated Bayesian inference. We further focus on dynamic smFISH experiments that perturb gene expression (e.g. induction) and then measure mRNA distributions at different times after induction to infer dynamics and kinetic parameters. Induction can change one or more of the model parameters (Figure 1c). The smFISH data from an induction experiment consist of a joint histogram $h = h(\Vec{\omega},t_\ell)$, where $t_\ell$ are independent observations made at different times before and after induction. If the changed parameters are unknown {\it a priori}, then one should evaluate all possible induction models, which leads to a combinatorial explosion in model space. For example, there are $2^8=256$ candidate induction models for the 3-state model shown in Figure 1c.

\begin{figure}[t]
\centering
\includegraphics[width=\linewidth]{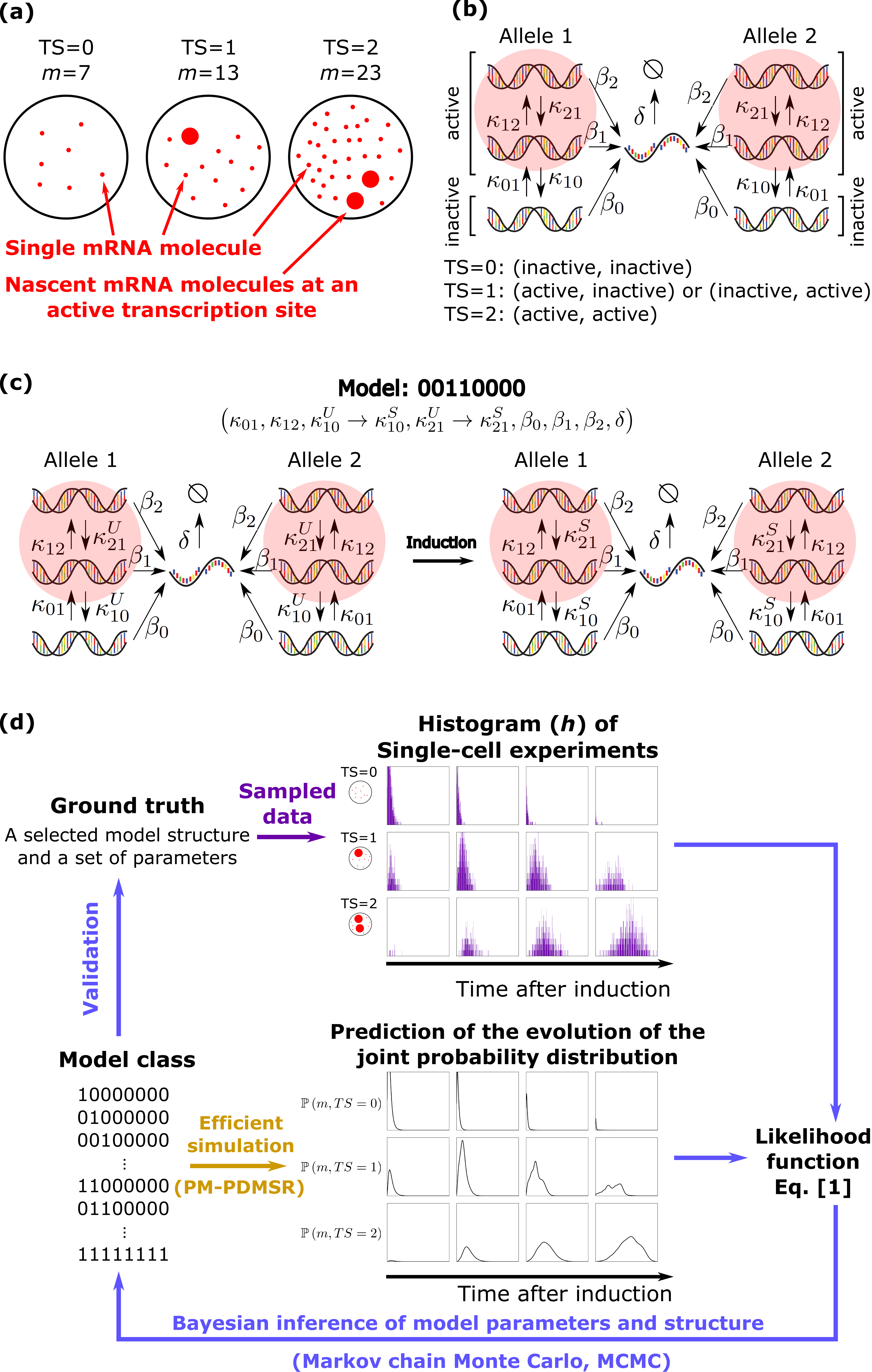}
\caption{Single cell data and models of gene expression. (a) The single-molecule RNA FISH (smFISH) technique provides information on the localization and numbers of mature mRNAs (m) in single cells, including clusters of nascent transcripts produced at transcription sites (TS) at “active” genetic loci. (b) A diploid, two-allele 3-state genetic model where $\kappa_{ij}$ is the transition rate between genetic states, $\beta_i$ is the mRNA synthesis rate of each state, and $\delta$ is the mRNA degradation rate. (c) Induction changes one or more parameters from an unstimulated (U) to a stimulated value (S). Here, we show one of many possible induction models $\mathcal{M}$, labelled in binary (00110000) (d) Schematic of the Bayesian inference work flow.}
\label{fig:1}
\end{figure}

A likelihood approach is used to connect mathematical models of stochastic gene expression to single cell data. Formally, the likelihood $\cal{L}$ is the probability that a candidate model $\mathcal{M}$ and its associated parameter set $\vec{\theta}$ would generate a given set of data ($h$). The number of parameters (i.e., dimension of $\vec{\theta}$) is determined by the model structure $\mathcal{M}$. Mathematically, the likelihood $\mathcal{L}$ is a function of the joint probability distribution $\mathbb{P}(\Vec{\omega},t_\ell \vert \vec{\theta},\cal{M})$ of a candidate model $ \mathcal{M}$ and its associated parameters $\vec{\theta}$ at discrete observation times:

\begin{align}
    \mathcal{L} ={}& \prod_{t_\ell \in \Phi} \bigg\{ \mathbb{M}_\ell \cdot \prod_{\vec{\omega} \in \Omega} [\mathbb{P}(\Vec{\omega},t_\ell | \Vec{\theta},\mathcal{M})]^{h(\Vec{\omega},t_\ell)} \bigg\} \label{eq:likelihoodDef}
\end{align}
where $\Phi$ is the set of observation times and $\mathbb{M}_\ell$ is the Multinomial coefficient associated with each $h(\Vec{\omega},t_\ell)$ that arises because the data were not ordered.  

In our Bayesian inference work flow (Figure 1d), each candidate model $\cal{M}$ in the class of possible models $\l\{\cal{M}\r\}$ will require a large number ($\ge 10^6$) of Monte Carlo steps where, at each step, numerical simulations calculate the time-dependent mRNA distributions and evaluate the likelihood that different parameter sets $\vec{\theta}$ for that model generated the observed data. Our previous software \cite{Gomez-Schiavon2017} took days to perform the likelihood calculations for one model, which highlights the challenge of using Bayesian inference to evaluate hundreds of models and perform model selection. Below, we develop a hybrid method that both accelerates numerical simulation and likelihood calculations, and (in contrast to standard methods) scales with the number of multi-core processors, thus, allowing for efficient parallelization. 

\subsection{A novel hybrid method to calculate the time evolution of discrete-state models}

While exact time-dependent solutions exist for two-state models \cite{Swain2008Analytical,Kumar2014Exact,Tiberi2018Bayesian}, it is hard to generalize this analysis to models with more than two states. It is therefore necessary to solve the general time-dependent problem using numerical simulations. There are two classes of numerical procedures to solve the time evolution of a discrete-state model for a given set of parameters.  The first class forward-evolves the chemical master equations (CME), which are a system of infinitely many coupled ordinary differential equations (ODEs) that describe the joint probability distributions $\mathbb{P}(\vec{\omega},t)$ as a function of time \cite{Munsky2006FSP,cao2016ACME}. To be numerically feasible, a truncation scheme (e.g. only consider mRNA levels below a maximum $M$) is used to reduce the infinite size of the dynamical system. While this approach delivers accurate estimates of the temporal evolution of the truncated system, there are two shortcomings. First, the number of ODEs scale as $S^2M$ where $S$ is the number of genetic states for each allele. The ODE system becomes unwieldy for mammalian cells where the number of observed mRNAs per cell can be $\mathcal{O}(10^3)$ \cite{raj2006,little2013,bahar-halpern2015,skinner2016}. Second, the forward integration of the CME requires stiff ODE solvers, which can place demands on memory resources and hinder parallel processing. The second class of numerical procedures utilize kinetic Monte Carlo methods (e.g., continuous time Markov chain simulation \cite{Gillespie76General,Gillespie77Exact}) to sample the temporal evolution of the joint probability distribution $\mathbb{P}(\vec{\omega},t)$. While this approach is computationally less expensive, it comes at the cost of having to sample over many runs to achieve equivalent accuracy to CME.

\begin{figure}[t]
\centering
\includegraphics[width=\linewidth]{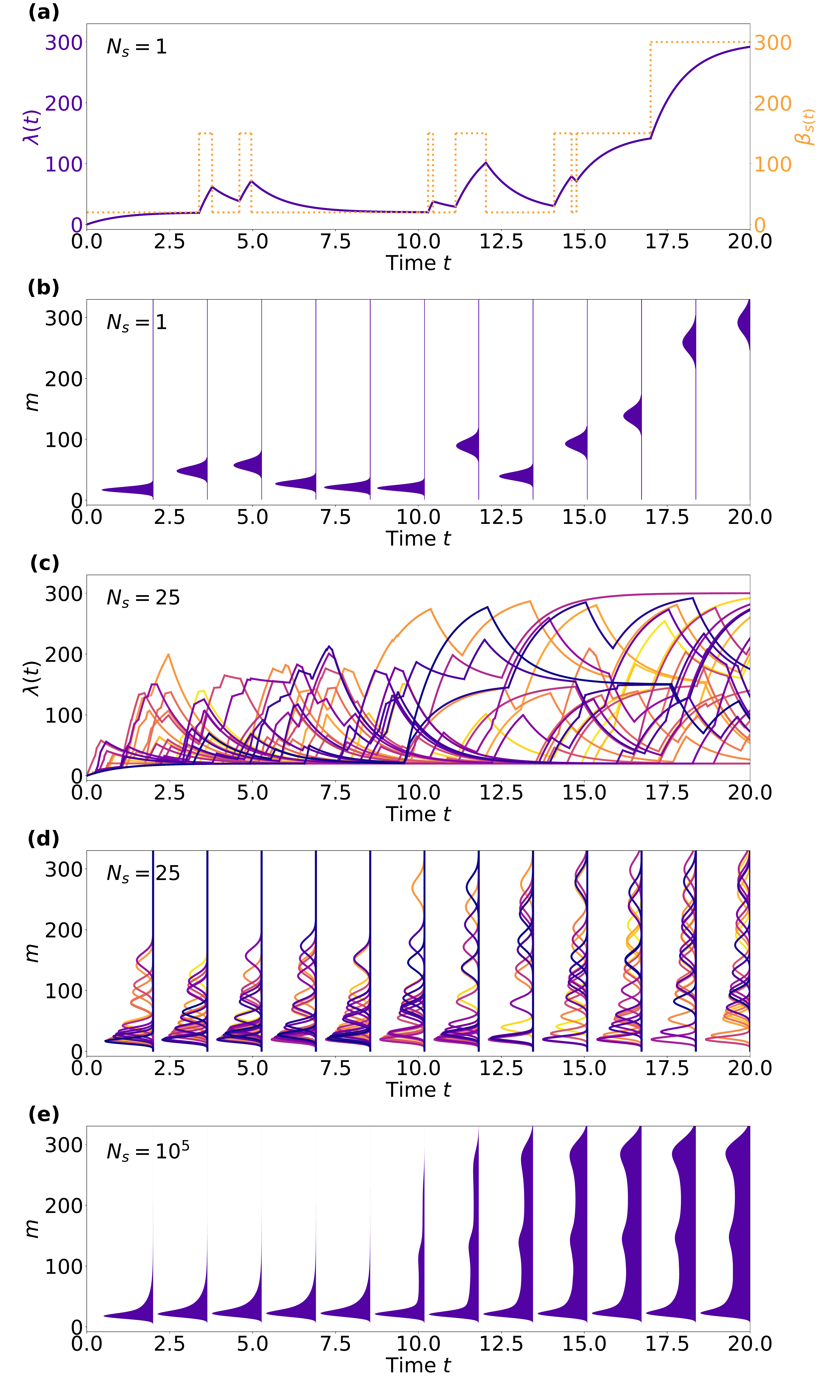}
\caption{Hybrid simulation method, PM-PDMSR. For simplicity, we illustrate the principle of PM-PDMSR for a single allele, 3-state model ($\mathcal{M}=00110000$). The gene is induced at $t=10$. Model parameters: before stimulation: $\l(\kappa_{01},\kappa_{12}, \kappa_{21}^{U}, \kappa_{10}^{U}, \beta_0, \beta_1, \beta_2, \delta \r) = \l(0.5, 0.5, 5, 5, 20, 150, 300,1\r)$; after stimulation, $\kappa_{21}^{S}=\kappa_{10}^{S} = 0.5$.(a,c) Changing transcription and dynamic rates for $N_s=1$ and $N_s=25$ sample paths. (b,d) Poisson mRNA distribution for the sample paths shown in (a,c), respectively. (e) Convolution of Poisson mixtures generated from $N_s=10^5$ sample paths.}\label{fig:2}
\end{figure}

In this article, we propose a hybrid simulation method (Poisson Mixture with Piecewise Deterministic Markov Switching Rates, or PM-PDMSR) which leverages analytical results and the efficiency of the kinetic Monte Carlo method. The key result is that the mRNA distribution can be exactly calculated for any realization (trajectory) of the genetic state, $s(t)$; see Methods. Once transient, initial conditions have burned off ($t \gg \delta^{-1}$), where $\delta$ is the mRNA degradation rate, the mRNA ($N_{\rm mRNA}$) distribution is always Poisson, $\mathbb{P}(N_{\rm mRNA}=m) = \lambda^m(t) e^{- \lambda(t)} / m!$ with a dynamic rate $\lambda(t)$ satisfying the following piecewise ODE: 
\begin{equation}
    \frac{\dd}{\dd t} \lambda(t) = \beta_{s(t)} - \delta \lambda(t) \label{eq:1}
\end{equation}
with an initial condition $\lambda(0)=0$. Given any trajectory $s(t)$, we can exactly compute the mRNA distribution $\mathbb{P}(m\vert s(t))$; see Figure 2a-b.  Our goal, however, is to determine the joint distribution $\mathbb{P}(\vec{\omega},t)$, which requires us to generate $N_s$ sample paths of $s(t)$ that cover $\mathbb{P}(s,t)$. The sample paths in the small genetic state space ($S^2$-dimensional) are efficiently generated using standard kinetic Monte Carlo methods. After accumulating a large number of sample paths $N_s$ generated by the underlying model, the mixture of the Poisson distributions recovers the mRNA distribution:
\eq{
\widehat{\mathbb{P}}\l(N_\text{mRNA}\l(t\r)=m,s\l(t\r)=i\r)= \sum_{k=1}^{N_s} \frac{\delta_{i,s_k\l(t\r)}}{N_s} \frac{\lambda_k^m\l(t\r) e^{-\lambda_k \l(t\r)}}{m!},
}{eq:convolutionMain}
where where $\lambda_k\l(t\r)$ is the solution of \eqref{eq:1} subject to the $k^\text{th}$ sample path of genetic switching trajectory $s_k\l(t\r)$ and $\delta_{i,j}$ is the Kronecker delta (see Figure 2c-e). 

A detailed description of the hybrid simulator is given in the Methods section. We evaluated the efficiency of the hybrid simulator relative to CME in performing a single step of the Bayesian inference work flow, i.e. simulate the joint distribution $\mathbb{P}(\vec{\omega},t)$ and calculate the likelihood $\cal{L}$ that this joint distribution produced a given data set ($h$). We bench-marked the simulators on diverse classes of discrete-state models, parameter sets, and data sets; see Figure 3. The hybrid simulator is up to $10^3$ more efficient for models with increased genetic states, $S=3$ and $4$. The efficiency gain of the hybrid simulator originates from the fact that $\mathbb{P}(m\vert s(t))$ is solved exactly in mRNA space (and is independent of the size of $M$) and that $\mathbb{P}(s,t)$ is sampled efficiently in genetic-state space via kinetic Monte Carlo techniques. Last, we tested the scaling of efficiency of different simulators on a modern machine, which can execute 32 parallel threads. The hybrid method scaled linearly with the number of processors, i.e. 32 processors in parallel ran just as fast as one processor.  Surprisingly, the simultaneous execution of 32 stiff ODE solvers for the CME simulator took 16 times longer per processor than one processor alone; see Methods. 

\begin{figure}
\centering
\includegraphics[width=\linewidth]{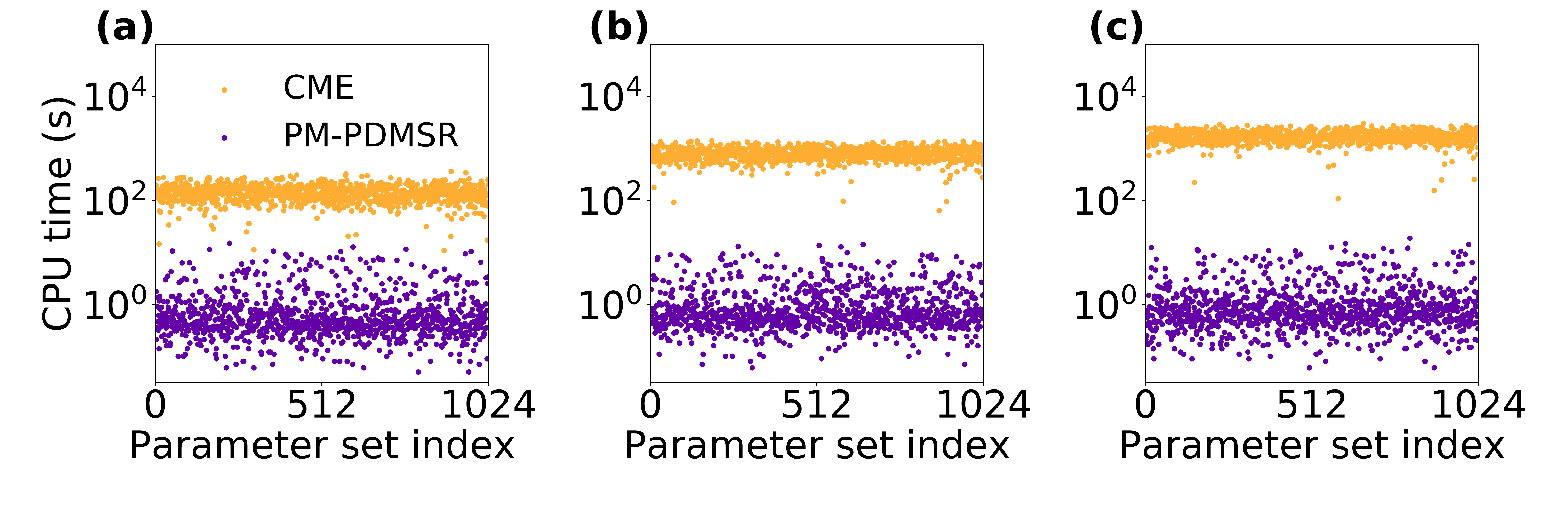}
\caption{Efficiency of the hybrid method relative to the CME method for a single processor. We measured the time for each method to complete one step of Bayesian inference, i.e. calculate joint distribution and evaluate the likelihood. This comparison was performed for increasingly complex model classes: (a) 2-state, (b) 3-state, (c) 4-state models of gene expression. Each model class was evaluated for 1024 different parameters along with associated data sets; see Methods for details.}
\label{fig:3}
\end{figure}

\subsection{Bayesian inference and uncertainty quantification of model parameters}

Equipped with an efficient simulator of the time-dependent joint probability distribution and likelihood calculation for any model and parameter set, we first turned our attention to uncertainty quantification of model parameters $\vec{\theta}$ for a fixed model $\cal{M}$. Given a likelihood, Bayesian inference uses Bayes formula to update any prior beliefs $\mathbb{P}(\Vec{\theta}|\cal{M}$) and calculate the posterior distribution $\mathbb{P}(\Vec{\theta}|h,\cal{M})$ of parameters $\vec{\theta}$ given the data $h$ and a fixed model $\cal{M}$: 

\begin{equation}
    \mathbb{P}(\Vec{\theta}\vert h,\mathcal{M}) = \frac{\mathbb{P}(h\vert\vec{\theta},\mathcal{M}) \mathbb{P}(\vec{\theta}\vert \mathcal{M})}{\mathbb{P}(h\vert \mathcal{M})} = \frac{\mathcal{L} \cdot \mathbb{P}(\vec{\theta}\vert \mathcal{M})}{\mathbb{P}(h\vert \mathcal{M})} \label{eq:MrBayesTalking}
\end{equation}

As done previously, we resorted to Markov chain Monte Carlo (MCMC) with a Metropolis--Hastings (MH) sampler to estimate the posterior distribution $\mathbb{P}(\Vec{\theta}|h,\cal{M})$; see Methods and \cite{Gomez-Schiavon2017}. We assumed that the prior $\mathbb{P}(\Vec{\theta}|\cal{M})$ is log-uniform. At each MCMC step, the MH sampler randomly proposes a nearby parameter set and computes the ratio of the posterior probability  $\mathbb{P}(\Vec{\theta}\vert h,\mathcal{M})$ relative to that of the current parameter set, and probabilistically accepts or rejects the proposal with a prescribed criterion that only depends on the ratio of the likelihood values. The denominator $\mathbb{P}(h\vert \mathcal{M})$ in \eqref{eq:MrBayesTalking} is identical for any parameter set $\vec{\theta}$ and cancels during the calculation of the ratio.

We bench-marked our approach on two synthetic data sets that were generated by sampling ($N=100$ or $1000$ cells per time point for a total of 4 time points) from a two-allele, 3-state induction model, where the induction stimulus decreased the downward transition rates; see Methods. Here, the model was known {\it a priori} and our goal was to infer the kinetic parameters and perform uncertainty quantification by comparing their posterior distributions (Figure 4a-b) to the ground truth (GT) parameters used to generate the sampled synthetic data set (Figure 4c-d). Our method constrained the posterior parameter distribution around the ground truth, and a 10-fold increase in the number of sampled cells dramatically reduced uncertainty in the inferred parameters. This observation holds true for a synthetic data set generated by a different two-allele, 3-state induction model; see Supplemental Materials.

\begin{figure*}
\centering
\includegraphics[width=\linewidth]{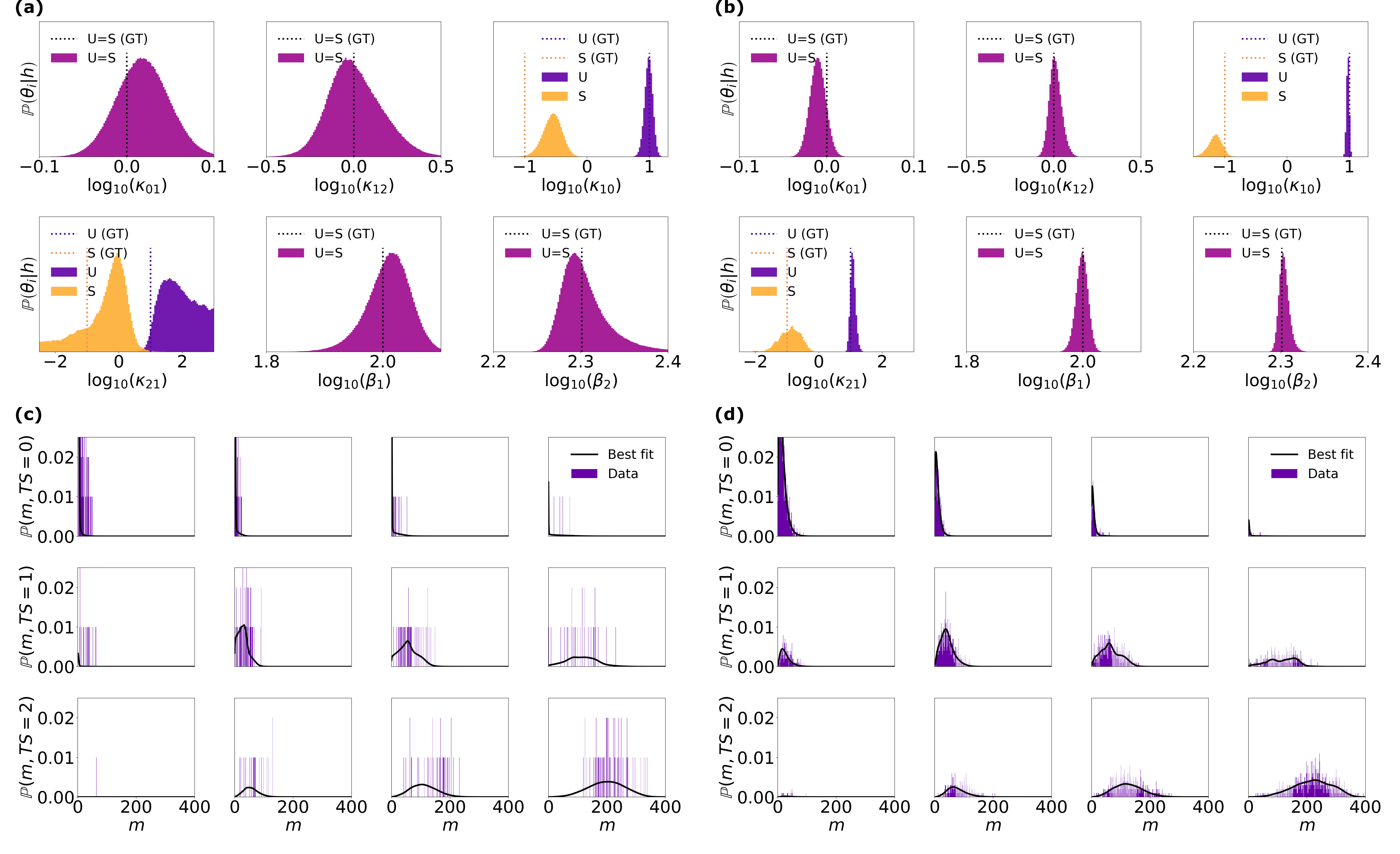}
\caption{Parameter inference and uncertainty quantification using the Bayesian posterior distribution. We bench-marked the hybrid method by running Bayesian inference on a synthetic data set sampled ($N$ cells at 4 different time points) from a known model $\mathcal{M}=00110000$ and ``ground-truth'' (GT) parameter set.  Posterior distributions (a-b) and joint distribution of best-fit parameters (c-d) for $N=100$ and $1000$ cells per time point, respectively.}
\label{fig:4}
\end{figure*}

\subsection{Model selection using the full Bayesian framework}

Knowing that our method of accelerated Bayesian inference can reliably constrain the kinetic parameters for a given model, we turned our attention to the harder problem of model selection. The goal was to identify the correct model from 64 possible types of two-allele, 3-state induction models given the same synthetic data set in Fig.~4, which was sampled from a ground-truth model and its parameters. Bayesian analysis naturally provides a quantitative measure of the likelihood of any model $\cal{M}$, i.e., the probability of the model to reproduce the experimentally observed data $h$.  The measure, referred to as the marginalized likelihood or {\it evidence} \cite{MacKay2005Information,sivia2006data}, is the denominator of \eqref{eq:MrBayesTalking}:
\begin{equation}
    \mathbb{P}(h\vert\mathcal{M}) = \int \underbrace{\mathbb{P}(h\vert\vec{\theta},\mathcal{M})}_{\mathcal{L}} \underbrace{\mathbb{P}(\vec{\theta}|\mathcal{M})}_{\text{Prior}} \dd \vec{\theta} \label{eq:evidenceDef}
\end{equation}
The evidence is simply the probability that a model $\cal{M}$ produced data $h$ and is equal to the sum of the probabilities of the model (i.e.~likelihood) over all sets of parameters that could have produced the data. The evidence is a convolution of the likelihood with the prior $\mathbb{P}(\vec{\theta} \vert\mathcal{M})$, which quantifies the belief regarding the initial parameter distributions. The dimensionality of $\vec{\theta}$ does not have to be identical for two different models and this prior inherently penalizes models with too many parameters; see Discussion. 

The evidence for a model $\mathcal{M}$ is not calculated during the MCMC sampling of the posterior distribution and has to be computed separately. Computing the evidence is a sophisticated problem \cite{Green1995Reversible,Kass1995BayesFactors,Weinberg2012Computing,Pajor2017Estimating} and we adopted an Importance Sampler of the Harmonic Mean Estimator (IS-HME) proposed Robert and Wraith \cite{Robert2009Computational}, which re-samples the posterior distribution estimated by the MCMC to compute the evidence of each model; see Methods. We first carried out the MCMC calculations of posterior distributions for each of the 64 possible types of two-allele, 3-state induction models for the data sets described in Fig.~4. We then used IS-HME to compute the evidence of each model given the underlying data set. We compared the IS-HME evidence to maximum likelihood metrics used for model selection, such as the Bayesian Information Criterion (BIC) and Akaike Information Criterion (AIC) \cite{Gomez-Schiavon2017}. Both BIC and AIC are approximations to the Bayesian evidence and become equivalent in the limit of large sample sizes; see Discussion. 

Our results demonstrate that the IS-HME evidence $\mathbb{P}(h\vert\mathcal{M})$ of the ground truth model dominates over other models ($\ge 95\%$) when using Bayesian inference on the larger data set ($N = 1000$ cells sampled per time point); see Figure 5. The BIC approximation also selected the ground-truth model (although incorrect models exhibited significant probabilities, e.g. $>5\%$) whereas the AIC failed to select the correct model. When the sample size dropped to $N=100$ cells per time point, even IS-HME evidence could not reliably select the ground-truth model with this under-powered data set. 
\begin{figure*}[t]
\centering
\includegraphics[width=.95\linewidth]{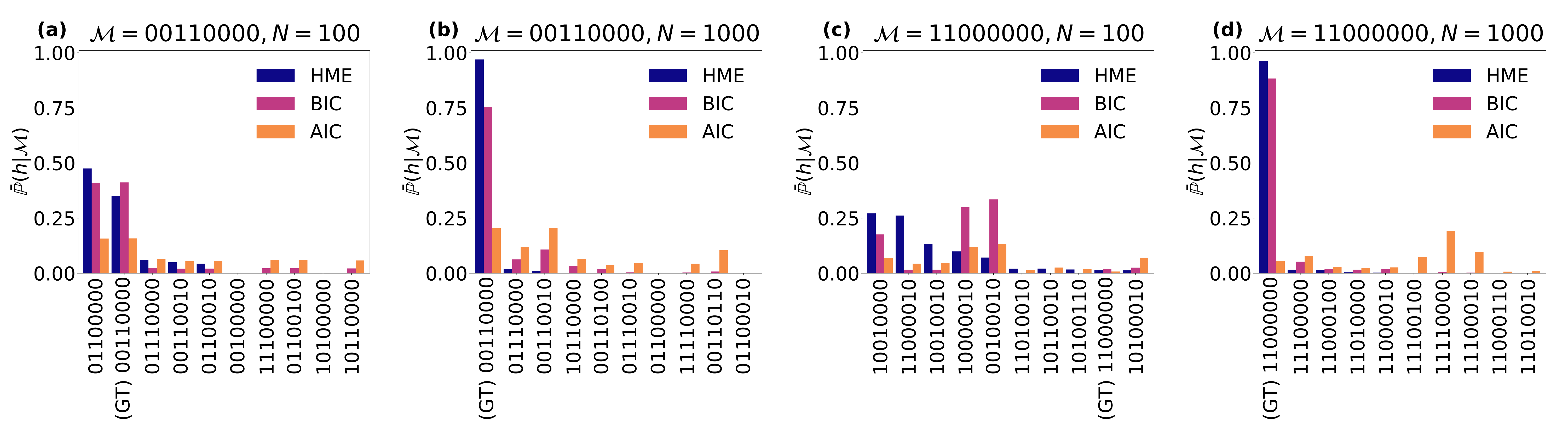}
\caption{Model selection using Bayesian evidence. We plot the IS-HME, BIC, and AIC evidence metrics of the top 10 models, ordered by decreasing IS-HME score. Model selection was performed on data sampled at two densities ($N$ = 100 and 1000 cells per time point at 4 different time points) for two different ground-truth models ($\mathcal{M}=0011000$ and $\mathcal{M}=1100000$).}
\label{fig:5}
\end{figure*}

\section{Discussion}

Piecewise-deterministic Markov processes (PDMP) have become a useful, coarse-grained description of stochastic gene dynamics, where the underlying discrete variable $s(t)$ captures the stochastic dynamics of gene states and the continuous variable $\lambda(t)$ captures the first moment of downstream gene products \cite{Bokes2013Transcriptional,lin2016gene,lin2016bursting,Hufton2017,Bressloff2017stochastic,Lin2018Stochastic,lin2018Efficient}. The key insight of our manuscript was proving that the time-dependent mRNA {\it distribution} of any underlying $s(t)$ is asymptotically a Poisson distribution with a rate $\lambda(t)$, and that the time-dependent joint probability distributions of discrete-state models are dynamic Poisson mixtures, whose mixing kernels are characterized by a PDMP. This significantly expands upon a related framework, which only considered the stationary distribution of discrete-state models \cite{Herbach2018}. More generally, our analysis helps bridge a gap between mechanistic discrete-state models and statistical models used in single cell analysis. For example, Wang et al. recently proposed an statistical model of gene expression, which postulated that mRNA distributions are Poisson mixtures \cite{Wang2018Gene}, and our work justifies this assumption. 

We used our insight to develop a hybrid method that calculates the time-dependent joint distribution more efficiently than standard numerical methods that forward-integrate the Chemical Master Equation (CME).  The efficiency arises because our method analytically solves the mRNA distribution and rapidly  samples many path $s(t)$ of discrete-switching events using a kinetic Monte Carlo algorithm.  We benchmarked the hybrid method and showed that it is $\mathcal{O}\l(10^3\r)$ more efficient than previous methods that directly integrate the CME. Furthermore, the hybrid method runs as efficiently in parallel on a multi-core processor than it does on a single processor. The stiff CME integrators ran more slowly in parallel and this sublinear scaling persisted for different integrators.  We suspect that the slow-down arises from the competing memory demands of stiff CME integrators running on a multi-core processor. While there is room to improve stiff integration and parallelization, current approaches are limited compared to the hybrid method because they must integrate the CME for a large number of mRNA states, e.g. 0 to 1000 mRNAs per cell.

We incorporated the hybrid algorithm into BayFISH and were able, for the first time, to use a full Bayesian framework for model selection and uncertainty quantification of parameters from single-cell smFISH data.  We adopted the Bayesian framework for model selection because it naturally quantifies ``Occam’s factor'' \cite{MacKay2005Information,Kass1995BayesFactors} and, thus, avoids over-fitting.  For example, the top models based on Bayesian evidence are not the most complex models with the largest number of parameters that change upon induction, e.g. $\mathcal{M}=11110110$; see Figure 5. The evidence resists over-fitting because when the dimensionality of parameter space is high, the value of a uniform probability density of the prior parameter distribution $\mathbb{P}\l(\vec{\theta}\vert \mathcal{M}\r)$ in \eqref{eq:evidenceDef} is small due to normalization. Thus, Bayesian evidence will favor a model that is complex enough to have a large likelihood but not too complex to decrease the prior parameter density. We note that when the data sample size is large, such that the posterior distributions $\mathbb{P}\l(\vec{\theta} \vert h, \mathcal{M}\r)$ can be approximated by a multivariate normal distribution, the logarithm of the evidence converges asymptotically to the Schwartz index (commonly known as the Bayesian Information Criterion, BIC) \cite{Kass1995BayesFactors,sivia2006data}. Here, we bench-marked the ability of Bayesian evidence and the BIC metric to select the correct model from synthetic data sets generated by a ground-truth model and parameters. Figure 5 shows that while the BIC (but not AIC) ranked models similarly to Bayesian evidence for the larger data set ($N$=1000 cells per time point), BIC requires an even larger sample size to confidently converge to the correct model. This is an important issue because most biology labs are ill-equipped to generate and analyze large smFISH data sets, and their sample sizes are typically $N$ = 100 - 1000 cells per time point. Our work demonstrates why Bayesian inference should be used for modestly sampled data sets.  We show that $N$=100 cells per time point is sufficient for parameter inference and uncertainty quantification if one has high confidence in the underlying model; see Figure 4.  However, if the goal of the smFISH experiments is model selection, then these smaller data sets are under-powered and the experimentalist needs to increase data sampling by atleast 10-fold; see Figure 5a and 5c. Here, we only considered one round of experiments followed by Bayesian inference, but multiple cycles of data collection and analysis are the norm.  Our framework quantifies certainty in both models and parameters using Bayesian evidence and posterior distributions. Future work can complete the data collection and analysis cycle by using the evidence and posterior distributions to rationally dictate the next round of experiments, i.e. different sampling times and densities, that are most informative for constraining models and/or parameters. 

In this article, we adopted a Markov chain Monte Carlo algorithm with Metropolis-Hastings sampling (i.e., MCMC-MH) to compute the posterior distributions of the model parameters \cite{Gomez-Schiavon2017,Munsky2018Distribution}. However, there is room to further improve the speed of Bayesian inference.  First, Hamiltonian Monte Carlo (HMC) algorithms are more efficient at sampling posterior distributions in high dimensional parameter spaces because they uses local sensitivity, i.e., the partial derivatives of the likelihoods with respect to the model parameters \cite{DUANE1987216,neal2011mcmc,betancourt2017conceptual}. Second, although PM-PDMRS is efficient at generating sample paths in the space $\l(s, \lambda\r)$, evaluating the convolution to calculate the joint distribution \eqref{eq:convolutionMain} is the rate-limiting step in the likelihood calculation. Thus, transforming the experimental data $h$ into the mixing kernel of the Poisson mixtures $\rho_s\l(\lambda\r)$ would accelerate Bayesian inference. Last, one could use low-order moments of PM-PDMSR and experimental data to formulate a sufficient statistics for likelihood-free Approximate Bayesian Computation \cite{Toni187Approximate}, thus, replacing the explicit calculation of the likelihood $\mathcal{L}$.

\section{Methods}
\subsection{Poisson mixture with piecewise deterministic Markov switching rates}

We illustrate our central theoretical results using a single-allele model. However, the results generalize to multiple-allele models because the states of a multiple-allele model can be relabeled as internal states of a single-allele model. \\

\newcommand{\Nini}{N_\text{mRNA}^\text{initial}}
\newcommand{\Nnew}{N_\text{mRNA}^\text{new}}
\newcommand{\Ntot}{N_\text{mRNA}}
\noindent{\bf Central theoretical result I:} Given a trajectory of genetic state $s(t)$, the total number of mRNA, $N_\text{mRNA}\left(t\right)$, is the sum of two variables $\Nini\left(t\right)$ and $\Nnew\left(t\right)$. $\Nini\left(t\right)$  describes the number of initial mRNAs that remain at time $t$. The probability distribution of $\Nini\l(t\r)$ is a Binomial mixture weighted by the initial mRNA distribution $\mathbb{P}_{m,0}:=\mathbb{P}\l(\Nini\l(t=0\r)=n\r)$:
\al{
\mathbb{P}\l(\Nini\l(t\r)=n\r) ={}& \sum_{m=0}^\infty \mathbb{P}_{m,0} \left(\begin{array}{c} m\\n \end{array}\right) \left(1-e^{-\delta t} \right)^{m-n} \times \nonumber \\ {}& e^{-n \delta t } \Theta\left(m-n+\frac{1}{2}\right).
}{}
Here, $\Theta\l(\cdot\r)$ is the Heaviside step function. $\Nnew\l(t\r)$ describes the number of new mRNAs that were synthesized after $t>0$ and still remain at time $t$. The probability distribution of $\Nnew\l(t\r)$ is a Poisson distribution with rate $\lambda\l(t\r)$
\eq{
\mathbb{P}\l(\Nnew \l(t\r) = m\r) = \frac{\lambda^m \l(t\r) e^{-\lambda\l(t\r)}}{m!},
}{eq:lambda}
where $\lambda(t)$ satisfies equation $\dot{\lambda}\l(t\r) = \beta_{s\l(t\r)}-\delta \lambda\l(t\r) $ with an initial condition $\lambda\l(0\r)=0$. \\

\noindent{\bf Proof.} We denote the probability of the total number of mRNA $\Ntot$ at time $t$ by $\mathbb{P}_m\l(t\r)$. For a given trajectory of the genetic state $s(t)$, the temporal evolution of   satisfies the chemical master equation (CME) 
\al{
\frac{\dd}{\dd t}\mathbb{P}_m\l(t\r) ={}& -\l(\beta_{s\l(t\r) + \delta m}\r)\mathbb{P}_m\l(t\r) \nonumber \\
{}& + \beta_{s\l(t\r)} \mathbb{P}_{m-1}\l(t\r) + \delta \l(m+1\r) \mathbb{P}_{m+1}\l(t\r),
}{eq:master}
for all $m\in \mathbb{Z}_{\ge 0}$ and with a boundary condition $\mathbb{P}_{-1}=0$. We prove the central theoretical result I by using the probability generating function defined by
\eq{
\mathcal{G}\l(z,t\r) := \sum_{m=0}^{\infty} z^m \mathbb{P}_m \l(t\r),
}{}
where $\mathbb{P}_m \l(t\r)= \partial_{z}^m \mathcal{G}\l(z,t\r)/m!$. We first solve for $\mathcal{G}\l(z,t\r)$ over an interval of time when $s\l(t\r)$ is constant. We will then extend our analysis to include piecewise intervals of time with different value of constant $s$, similar to the $s\l(t\r)$ generated by a genetic state model. To begin, we apply the operator $\partial_t$ to $\mathcal{G}\l(z,t\r)$ and use \eqref{eq:master} to obtain the partial differential equation (PDE):
\eq{
\partial_t \mathcal{G}\l(z,t\r) = \delta \l(1-z\r) \partial_z\mathcal{G}\l(z,t\r) + \beta_s \l(z-1\r) \mathcal{G}\l(z,t\r).
}{}
This linear PDE can be solved using the method of characteristics \cite{vankampen2007} and the general solution is
\al{
\mathcal{G}\l(z,t\r) ={}& \l[\sum_{m=0}^\infty \l(1+ \l(z-1\r) e^{-\delta t}\r)^m \mathbb{P}_{m,0}\r] \times  \nonumber \\
{}& \exp \l(\frac{\beta_s }{\delta} \l(z-1\r) \l(1-e^{-\delta t}\r)\r),
}{eq:constSsol}
where $\mathbb{P}_{m,0}:=\mathbb{P}_{m,0}\l(t=0\r)$ is the initial mRNA distribution of the system. For reasons that will become apparent below, we label the initial-distribution-dependent part by $\mathcal{G}^\text{init}\l(z,t\r)$ and the rest of the terms by $\mathcal{G}^\text{new}\l(z,t\r)$:
\subeq{
\mathcal{G}^\text{init}\l(z,t\r)={}& \sum_{m=0}^\infty \l(1+ \l(z-1\r) e^{-\delta t}\r)^m \mathbb{P}_{m,0}, \\
\mathcal{G}^\text{new}\l(z,t\r)={}&\exp \l(\frac{\beta_s }{\delta} \l(z-1\r) \l(1-e^{-\delta t}\r)\r).
}{}
The above solution applies to a constant $s\l(t\r)$. We now consider a piecewise-constant trajectory for any genetic state $s\l(t\r)$. To specify the discrete state and the switching events, we label $s\l(t\r)$ by the ordered pairs $\l(t_i,s_i\r)$ for $i=0,1,\ldots,N$ before an observation time $t$. The gene starts with a state $s_0$ at $t_0:=0$, switches to $s_1$ at $t_1$,\ldots until the final switching event to $s_N$ at time $t_N$. Our aim is to compute the generating function $\mathcal{G}\l(z,t\r)$ after $N$ switching events ($t\ge t_N$). 

The solution $\mathcal{G}\l(z,t,\vert t\le t_1\r)$ is identical to \eqref{eq:constSsol} with $s=s_0$ before the first switching event. At $t=t_1$, the generating function is
\eq{
\mathcal{G}\l(z,t_1\r)=\mathcal{G}^\text{init}\l(z,t_1\r) \times \mathcal{G}^\text{new}\l(z,t_1\r).
}{}
Note that after $t_1$ and before $t_2$, the genetic state changes to $s_1$ and only the transcription rate in \eqref{eq:master} changes from $\beta_{s_0}$ to $\beta_{s_1}$. The initial condition of the generating function of this period ($t_1\le t \le t_2$ ) is precisely $\mathcal{G}\l(z,t_1\r)$ in the above equation. Matching the ``initial condition'' for $\mathcal{G}\l(z,t_1\r)$, we arrive at
\al{
{}&\mathcal{G}\l(z,t\vert t_1 \le t \le t_2 \r) =\mathcal{G}^\text{init}\l(z,t\r)  \cdot \exp\l\{\frac{z-1}{\delta}  \l[\vphantom{\l(1-e^{-\delta \l(t-t_1\r)}\r)}{\beta_{s_0}} \times \r.\r. \nonumber \\
{}&\l.\vphantom{\frac{z-1}{\delta}} \l.\l(1-e^{-\delta t_1}\r) e^{-\delta \l(t-t_1\r)} + {\beta_{s_1}}\l(1-e^{-\delta \l(t-t_1\r)}\r)\r]\r\}.
}{}
We iterate this procedure for each piecewise ``episode'' until $t=t_N$,
\al{
\mathcal{G}{}&\l(z,t_N\r) =\mathcal{G}^\text{init}\l(z,t_N\r) \times \nonumber \\
{}& \exp\l[\frac{z-1}{\delta}\sum_{n=1}^{N} \beta_{s_{n-1}} \l(1-e^{-\delta\l(t_n - t_{n-1}\r)}\r)e^{-\delta\l(t_N - t_n\r)}\r],
}{}
and for $t\ge t_N$, 
\al{
\mathcal{G}\l(z,t\vert t\ge t_N\r)={}& \mathcal{G}^\text{init}\l(z,t\r) \times \mathcal{G}^\text{new}\l(z,t\r), \nonumber\\
\mathcal{G}^\text{new}\l(z,t\r):={}& \exp\l[\psi\l(\l\{t_n\r\}_{n=1}^N\r)\r], \nonumber \\
\psi\l(\l\{t_n\r\}_{n=1}^N\r):={}& \frac{z-1}{\delta} \l[ \sum_{n=1}^{N} \beta_{s_{n-1}} \l(1-e^{-\delta\l(t_n - t_{n-1}\r)}\r)\times \r. \nonumber \\
{}&\l.\vphantom{\sum_{n=1}^{N}} e^{-\delta\l(t-t_N\r)}+ \beta_{s_N}\l(1-e^{-\delta \l(t-t_N\r)}\r) \r]
}{eq:finalSol}
The total solution $\mathcal{G}\l(z,t\vert t\ge t_N\r)$ is factorized into two terms, $\mathcal{G}^\text{init}\l(z,t\r)$ and $\mathcal{G}^\text{new}\l(z,t\r)$, for any $N$ and $t$. The probability generating function of the sum of two independent random variables is the product of the generating functions of the random variables. This hints that we can define two random variables, $X_1$ and $X_2$, which have generating functions $\mathcal{G}^\text{init}\l(z,t\r)$ and $\mathcal{G}^\text{new}\l(z,t\r)$ respectively. 

Our next task is to identify the variables $X_1$ and $X_2$ and their probability distributions. For $X_1$, we expand $\mathcal{G}^\text{init}\l(z,t\r)$ to arrive at
\eq{
\mathcal{G}^\text{init}\l(z,t\r) \equiv \sum_{m=0}^\infty \mathbb{P}_{m,0} \sum_{n=0}^\infty \l(\begin{array}{c}m\\n\end{array}\r) \l(1-e^{-\delta t}\r)^{m-n} z^n e^{-n\delta t}.
}{}
Recall that the generating function of a $\text{Binomial}\l(m,p\r)$ distribution is 
\eq{
\l[1-p + pz\r]^m = \sum_{n=0}^{\infty } \l(\begin{array}{c}m\\n\end{array}\r) \l(1-p\r)^{m-n} z^n p^n.
}{}
The probability distribution of $X_1$ is therefore identified to be a binomial mixure with a temporally decaying parameter $p=\exp\l(-\delta t\r)$ and a mixing kernel defined by the initial distribution $\mathbb{P}_{m,0}$. The physical meaning of $X_1\l(t\r)$ is the number of initial mRNA molecules that remain at time $t$, i.e., $\Nini\l(t\r)$. These mRNA can only degrade with the decay rate $\delta$. Each of the mRNA decays independently and, at time $t$, there is a probability $\exp\l(-\delta t\r)$ that a specific mRNA has not degraded. Importantly, when $t\gg 1/\delta$, this distribution will be concentrated at $m=0$ (see Corollary I below). 

The total mRNA $N\l(t\r)=X_1\l(t\r)+X_2 \l(t\r)= \Nini\l(t\r) + X_2\l(t\r)$, so $X_2\l(t\r)$ is identified to be the number of new mRNA molecules that were synthesized \emph{after} $t=0$ but which have not degraded at time $t$. We refer to this variable as $\Nnew\l(t\r)$. The squarer bracket of $\mathcal{G}^\text{new}\l(z,t\r)$ in \eqref{eq:finalSol} is the piecewise solution $\lambda\l(t\r)$ of the following ODE for a given genetic trajectory $s\l(t\r)$:
\eq{
\frac{\dd}{\dd t} \lambda\l(t\r) = \beta_{s\l(t\r)}-\delta \lambda\l(t\r), \quad \text{and } \lambda\l(0\r) = 0. 
}{eq:ode}
We now expand $\mathcal{G}^\text{new}\l(z,t\r)$:
\al{
\mathcal{G}^\text{new}\l(z,t\r)={}&\exp\l[\l(z-1\r) \lambda\l(t\r)\r] = e^{-\lambda\l(t\r)} \sum_{m=0}^\infty \frac{z^m \lambda^m \l(t\r)}{m!}\nonumber \\
= {}& \sum_{m=0}^\infty z^m \frac{e^{-\lambda\l(t\r)} \lambda^m \l(t\r)}{m!} = \sum_{m=0}^\infty z^m q_m \l(\lambda \l(t\r)\r),
}{}
where $q_m\l(\lambda\l(t\r)\r)$ is the probability density function of a Poisson distribution with rate $\lambda\l(t\r)$, as in \eqref{eq:lambda}.$\ \square$\\

\noindent{\bf Corollary I.} The transient timescale for the initial distribution is $\mathcal{O} \l(1/\delta\r)$. When $t\gg \mathcal{O}\l(1/\delta\r)$, the mRNA distribution converges to a Poisson with a dynamic rate parameter $\lambda\l(t\r)$. \\

\noindent{\bf Proof.} Physically, the timescale of degradation of each initially populated mRNA is $1/\delta$, so for a timescale which is much longer than this, the initial distribution will be fully degraded. Mathematically, the probability that the initial mRNA molecules have not fully decayed is 
\al{
\mathbb{P} \l(\Nini\l(t\r) >0 \r) ={}& 1 - \mathbb{P}\l(\Nini\l(t\r)=0\r) \nonumber \\
{}&1- \sum_{m=1}^{\infty} \mathbb{P}_{m,0} \l(1-e^{-\delta t}\r)^m. 
}{}
In the asymptotic limit $t\gg 1/\delta$, $\exp\l(-\delta t\r)\ll 1$ so 
\al{
{}& \mathbb{P} \l(\Nini\l(t\r) >0 \r) \nonumber \\={}& 1- \sum_{m=0}^\infty \mathbb{P}_{m,0} \l(1-m e^{-\delta t}\r)\l[1+\mathcal{O} \l(e^{-\delta t}\r)\r] \nonumber \\
={}& \l\langle \Nini\l(0\r) \r\rangle e^{-\delta t} \l[1+\mathcal{O} \l(e^{-\delta t}\r)\r].
}{}
Here, $\l\langle \Nini\l(0\r) \r\rangle$ is the first moment of the initial distribution. $\mathbb{P} \l(\Nini\l(t\r) >0 \r)$ decays exponentially fast, and we can bound this probability to be smaller than $\varepsilon$ when $t > \delta^{-1} \log \l(\l\langle \Nini\l(0\r) \r\rangle/\varepsilon\r)$. $\ \square$ \\

For our numerical simulations detailed below, our total time was $20/\delta$. The contribution of the initial distribution $\mathbb{P} \l(\Nini\l(t\r) >0 \r) =\l\langle \Nini\l(0\r) \r\rangle e^{-20} \approx 2\times10^{-9} \times \l\langle \Nini\l(0\r) \r\rangle \approx 10^{-6}$, because typical numbers of measured mRNAs are bounded by $\mathcal{O} \l(500\r)$. \\

\noindent{\bf Central theoretical result II:} At long times $t \gg 1/\delta$, the mRNA distribution asymptotically converges to a Poisson mixture regardless of the initial mRNA distribution and genetic switching trajectory $s\l(t\r)$
\eq{
\mathbb{P} \l(N_\text{mRNA}\l(t\r)=m, s\l(t\r)=i\r) = \int_{0}^\infty \rho_i \l(\lambda,t\r) \frac{\lambda^m e^{-\lambda}}{m!} \dd \lambda,
}{eq:convoluted}
where the joint probability density $\rho_i\l(\lambda,t\r)$ satisfies the forward Kolmogorov equation
\eq{
\partial_t \rho_i = - \partial_\lambda \l[\l(\beta_i - \delta \lambda \r) \rho_i\r] + \sum_{j\ne i} \l(\kappa_{ji} \rho_j - \kappa_{ij} \rho_i\r).
}{eq:PDMPjoint}
The initial condition for $\rho_i \l(\lambda, t=0\r)$ are defined by
\eq{
\rho_i \l(\lambda, t=0\r):= \delta\l(\lambda\r) \mathbb{P}\l(s\l(t=0\r) =i \r),
}{}
where $\delta\l(\lambda\r)$ is the Dirac delta distribution at $\lambda=0$. \\

\noindent{\bf Proof.} The solution of \eqref{eq:ode} subject to random switching events of $s\l(t\r)$ is a random process. Formally, $\lambda\l(t\r)$ \emph{and} the discrete switching states $s\l(t\r)$ jointly comprise a piecewise deterministic Markov process (PDMP) \cite{Davis1983Piecewise,Bena2006Dichotomous,Faggionato2009NonEquilibrium}. The forward Kolmogorov equation describing the temporal evolution of the joint probability distribution is \eqref{eq:PDMPjoint} \cite{Hufton2017,Lin2018Stochastic}. Therefore,
\al{
\mathbb{P} {}&\l(N_\text{mRNA}\l(t\r)=m, s\l(t\r)=i\r) = \nonumber \\
{}&\int_{0}^\infty \mathbb{P}\l(N_\text{mRNA}\l(t\r) =m \vert \lambda\l(t\r)=\ell, s\l(t\r)=i \r)\rho_i \l(\ell,t\r) \dd \ell,
}{}
We then use the central theoretical result I and corollary I to show that $\mathbb{P}\l(N_\text{mRNA}\l(t\r) =m \vert \lambda\l(t\r)=\ell, s\l(t\r)=i \r) = \ell^m e^{-\ell}/m! $ asymptotically when $t\gg 1/\delta$ to complete the proof. $\ \square$\\

\noindent{\bf Efficient numerical method for sampling $\rho_i\l(\lambda,t\r)$.} Because $\lambda$ is continuous, solving the forward Kolmogorov equation \eqref{eq:PDMPjoint} is as complex as solving the full CME, both of which are infinite dimensional systems. Instead, we used an efficient kinetic Monte Carlo simulation \cite{Lin2018Stochastic,Bokes2013Transcriptional} to generate a large number of sample paths to estimate the asymptotic joint distribution $\mathbb{P}\l(N_\text{mRNA}\l(t\r) =m \vert \lambda\l(t\r)=\ell, s\l(t\r)=i \r)$ when $t\gg 1/\delta$ using \eqref{eq:convoluted}. The pseudo code of this algorithm is shown in Algorithm 1. \\

\begin{figure*}
\begin{algorithmic}[1]
\Require Initial state $\lambda\l(t=0\r)=0$ and $s\l(t\r)=s_0$. Kinetic rate $\kappa_{ij}$ (switching rates from discrete state $i\rightarrow j$), $\beta_k$ (transcription rates), and $\delta$ (degradation rate). $N$ discrete observation times $\mathcal{T}:=\l\{t_\ell\r\}_{\ell=1}^N$.
\Ensure {An exact sample path of the random process $(\lambda\l(t\r), s\l(t\r))$  at $N$ discrete times $\mathcal{T}$.} 
\State{$t \gets 0$, $\lambda \gets 0$, $s \gets s_0$} \Comment{Initiate system time and state}
\For{$t_\text{observation}$ in $\mathcal{T}$}
\While{$t<t_\text{observation}$}
\State{$\kappa \gets \sum_{i} \kappa_{si}$} \Comment{Compute the total propensity of switching}
\State{$u \gets \text{Unif}\l(0,1\r)$}
\State{$\Delta t \gets - \kappa^{-1} \log \l(u\r)$} \Comment{Sample the random advanced time}
\If{$t+\Delta_t < t_\text{observation}$}\Comment{A switching event occurs before $t_\text{observation}$}
\State{$c_0 \gets 0$, $c_i\gets \sum_{j-1}^i \kappa_{sj}$ for $i \in \l\{1,2,\ldots S\r\}$}  \Comment{Sample the switching events}
\State{$k\gets 0$}
\State{$w\gets c_S \times \text{Unif}\l(0,1\r)$}
\While{$w>\kappa_{c_k}$}
\State{$k \gets k +1$}
\EndWhile
\State{$\lambda \gets \beta_s/\delta + \l(\lambda - \beta_s /\delta \r) \exp\l(-\delta \Delta t\r)$, $s\gets k$} \Comment{Update system state}
\Else	\Comment{No switching event occurs before $t_\text{observation}$}
\State{$\Delta t \gets t_\text{observation} - t$}
\State{$\lambda \gets \beta_s/\delta + \l(\lambda - \beta_s /\delta \r) \exp\l(-\delta \Delta t\r)$}  \Comment{Update system state}
\EndIf
\State{$t \gets t+\Delta t$} \Comment{Update system time}
\EndWhile
\State{Output the system state $\l(\lambda,s\r)$ at the observation time $t_\text{observation}$}
\EndFor
\end{algorithmic}
\begin{flushleft}
Algorithm I: An efficient kinetic Monte Carlo algorithm which generates exact sample paths of the piecewise deterministic Markov process $\l(\lambda\l(t\r), s\l(t\r)\r)$.
\end{flushleft}
\end{figure*}

\noindent{\bf Computing the joint distribution from sample paths.} To compute the time-dependent joint distribution of genetic states and mRNAs, we generated $N_s$ sample paths $\l\{\lambda_k \l(t\r), s_k\l(t\r) \r\}_{k=1}^{N_s}$ with Algorithm I.  We then used \eqref{eq:convoluted} to estimate $\rho_{i} \l(\lambda\r)$ 
\eq{
\widehat{\mathbb{P}}\l(N_\text{mRNA}\l(t\r)=m,s\l(t\r)=i\r)= \sum_{k=1}^{N_s} \frac{\delta_{i,s_k\l(t\r)}}{N_s} \frac{\lambda_k^m\l(t\r) e^{-\lambda_k \l(t\r)}}{m!},
}{}
where $\delta_{i,j}$ is the Kronecker delta function which is equal to $1$ if $i=j$, and $0$ otherwise. We determined that $N_s\equiv 10^5$ is a sufficient number of sample paths to estimate the same joint distribution obtained by forward-integrating the CME. We refer to our method as the Poisson Mixture with a Piecewise Deterministic Markov Switching Rate (PM-PDMSR).

The goal was to compute the joint distribution of genetic states and mRNAs before and after induction. Similar to the situation in many experimental systems \cite{neuert2013,senecal2014}, the joint distribution is at stationarity before induction. Upon induction, we assume some model parameters are changed, which results in the time-evolution of the joint probability distribution $\mathbb{P}\l(N_\text{mRNA}\l(t\r)=m,s\l(t\r)=i\r)$ towards a new stationary state. We label the kinetic rates ($\kappa_{ij}$ and $\beta_k$) before and after induction by $U$ (Unstimulated) and $S$ (Stimulated). To use PM-PDMSR to estimate the stationary distribution before induction, we first solved for the marginal stationary distribution of the genetic state $p_{i}^\ast$, where $0=\sum_j \l(\kappa_{ij}^U - \kappa_{ji}^U\r)p_{i}^\ast$, $i=1,2,\ldots, S$  for an $S$-state model. We initiated $N_s p^\ast_i$ sample paths in PM-PDMSR at $\lambda = \beta_i/\delta$ and state $s=i$ and ran for $t=10/\delta$ so that the Poisson mixture relaxes to stationarity. Upon induction at $t=10/\delta$, we changed model parameters $\kappa_{ij}^U \rightarrow \kappa_{ij}^S$ and $\beta_k^U \rightarrow \beta_k^S$, for $i,j,k \in \l\{1,2,\ldots S\r\}$ and continued simulating the temporal evolution of the joint probability distribution after induction using PM-PDMSR. This is valid because our previous proofs and arguments for \eqref{eq:convoluted} apply even when the kinetic rate constants change upon induction. 

\subsection{Testing the speed and accuracy of the simulators}
We bench-marked the efficiency of PM-PDMSR versus the ``gold-standard'' simulator, i.e., forward-integration of a truncated CME \cite{Munsky2006FSP,cao2016ACME}. Both simulators were embedded into BayFISH and evaluated on their ability to perform a single Monte Carlo step, i.e., simulate the time-dependent joint distribution of a model and its parameters and to calculate the likelihood of a synthetic data set generated by the same model and parameters. This single-step bench-marking was performed for 1024 diverse models and parameters across two-allele, discrete-state models of increasing complexity (2-state, 3-state, and 4-state induction models).\\

\noindent{\bf Generating diverse models, parameters, and synthetic data sets.} Each model had a randomly chosen subset of parameters that were one value $\l(\kappa_{ij}^U, \beta_k^{U}\r)$ before induction from $t=0$ to $t=20$ and with different parameters $\l(\kappa_{ij}^S, \beta_k^{S}\r)$ after induction from $t=20$ to $t=22$. The genetic switching rates of parameters $\kappa_{ij}$ ranged between $10^{-2}$ to $10^2$ and the transcription rates $\beta_k$ range between $0$ and $200$. For each instance of a model, we randomly generated the switching rate constants in the logarithmic space ($log_{10} \kappa_{ij} \sim \text{Unif}\l(-2,2\r)$ if $\vert i - j \vert=1$) and transcription rate constants in linear space ($\beta_k \sim \text{Unif}\l(0,200\r)$). Note that not all parameters were random or changed upon induction: We fixed $\beta_0=0$ and the mRNA degradation rate $\delta=1$, and we constrained $\beta_i \le \beta_j$ if $i<j$. For the purpose of bench-marking algorithm speed and efficiency, we considered the most complex induction model $\mathcal{M}$ for each discrete model class. The corresponding induction model is $\mathcal{M}=11010$ (2-state), $\mathcal{M}=11110110$ (3-state), and $\mathcal{M}=11111101110$ (4-state).

The synthetic data of each model with its parameters were generated by running 1000 independent trajectories of standard continuous-time Markov chain simulation. We collected the statistics of the trajectories at four discrete times $t_\ell=20$, $20.5$, $21$, and $22$ ($N=1000$ cells per time point). The measured allele activity state $TS$ was marginalized: we define $TS=1$ when its internal state $s>0$. The synthetic data set therefore consists of a histogram at discrete times $t_\ell$, $h\l(m, TS, t_\ell\r)$, which is the number of trajectories with $m$ mRNAs and $TS$ active transcription sites (which can be $0$, $1$, or $2$ for a two-allele system) at time $t_\ell$. For each model class (2-, 3-, and 4-state models), we repeated the process 1024 times to test diverse parameter combinations. \\

\noindent{\bf CME simulators.} Given an induction model and associated parameters, we forward propagated the CME using the same parameter values used to create the synthetic data sets in the previous section.  We truncated the number of mRNA at $500$ (i.e., there are no transcription event once the system reaches $N_\text{mRNA}=500$)   with an absolute error tolerance of $10^{-5}$. The truncation number $M$ was motivated by data sets in animals, showing that mRNA populations can be as large as $\mathcal{O}\l(500\r)$  \cite{raj2006,little2013,bahar-halpern2015,skinner2016}. We tested different software platforms, including Matlab, Python (SciPy), and a research software ACME \cite{cao2016ACME} to forward integrate the same stiff CME. Python’s stiff integrator (using backward differentiation formula, BDF) outperformed other integrators and software platforms. Thus, Python (with SciPy) was chosen to be the platform for direct integration of the CME in the following analysis.  \\

\noindent{\bf Comparison of the PM-PDMSR and CME simulators.} 
We incorporated these PM-PDMSR and CME simulators into modified BayFISH software to evaluate their speed and accuracy for one Monte Carlo step. Both algorithms were implemented in c++ and compiled using Intel’s icc compiler. All PM-PDMSR and CME simulations were carried out on the same machine with Intel(R) Xeon(R) CPU E5-2695 v3 @ 2.30GHz. We computed the joint distributions of the models and parameter sets used to create the synthetic data sets. These joint distributions and corresponding synthetic data ($h\l(m, TS, t_\ell\r)$) were then used to compute the likelihood $\mathcal{L}$ of the generated data $h\l(m, n_{TS}, t_\ell\r)$ using \eqref{eq:likelihoodDef}. The execution time of a Monte Carlo step for each simulator for each model class of the generated parameter set was recorded and presented in Fig.~\ref{fig:3}. We also compared the accuracy of the calculated likelihoods of each synthetic data set. We compared the average error of the PM-PDMSR likelihood relative to the more accurate CME likelihood. We define the average error by
\eq{
\l\langle \varepsilon \r\rangle := \l\langle \l\vert \frac{\mathcal{L}_\text{PM-PDMRS} - \mathcal{L}_\text{CME}}{\mathcal{L}_\text{CME}} \r\vert \r\rangle
}{}
The relative accuracy of CME versus PM-PDMSR is presented in Supplemental Materials.

To test the parallelization of each simulator, we simultaneously ran 32 simulations on a 32-core machine (Intel(R) Xeon(R) CPU E5-2695 v4 @ 2.10GHz) and recorded the execution time; see Fig.~\ref{fig:3} PM-PDMRS on 32 parallel threads takes the same amount of time as running a single thread on a 32-core machine. In contrast, CME on 32 parallel threads takes 16 times longer than a single thread on a 32-core machine. This sub-optimal scaling holds true on the multiple machines that we tested and our algorithms leverage Python’s sub-processes functionality (Pool). We suspect that the slow down in CME is due to the high memory demand of the BDF integrator. 

\subsection{Bayesian statistical inference for model parameters and structure} 

\noindent{\bf Synthetic data.} We synthesized data sets to test if Bayesian statistical inference could identify the ground truth (of the model parameter values and the model structure.) We chose two 3-state ground-truth models for data synthesis: (1) a ON-rate induction model, $\kappa_{01}^U=1 \rightarrow \kappa_{01}^S = 12$, $\kappa_{12}^U=0.25 \rightarrow \kappa_{12}^S=20$, $\kappa_{10}^U=\kappa_{10}^S=3$, $\kappa_{21}^U = \kappa_{21}^S = 10$, $\beta_0=0$, $\beta_1=25$, and $\beta_2=300$, and (2) an OFF-rate induction model,  $\kappa_{01}^U=\kappa_{01}^S = 1$, $\kappa_{12}^U=\kappa_{12}^S=1$, $\kappa_{10}^U=10\rightarrow \kappa_{10}^S=0.1$, $\kappa_{21}^U =10 \rightarrow \kappa_{21}^S = 0.1$, $\beta_0=0$, $\beta_1=100$, and $\beta_2=200$. The protein degradation rate constant is $\delta=1$ by choosing the timescale of the model. We relaxed the models from $t=0$ to $t=20$, and sampled the system at  $t=20$, $20.5$, $21$, and $22$. The observation timescale was motivated by our recent experimental procedure \cite{Gomez-Schiavon2017}. For each model, we synthesized by sampling $100$ and $1000$ synthetic data at each of the descrete sampled times from the joint probability distribution. The data consists of the sampled and discrete number of mRNA, and whether the gene is active. Again, the genetic space $s$ is marginalized that we defined $s>0$ is an active allele ($TS=1$), otherwise inactive ($TS=0$).   \\

\noindent{\bf Bayesian analysis.} The model class we considered for Bayesian inference is the set of two-allele, 3-state models with $\beta_0^U=\beta_0^S=0$. The rest of the parameters are free parameters, but depending on the model structure, some of the perturbed ($S$) parameters may be constrained to the unperturbed ($U$) value. Combinatorially, there are in total $2^6=64$ models we considered, as there are six biophysical parameters $\vec{\theta} :=\l(\kappa_{01},\kappa_{12},\kappa_{21},\kappa_{10},\beta_1,\beta_2\r)$. 

We adopted a plain Markov chain Monte Carlo algorithm to sample the posterior distribution $\mathbb{P}\l(\vec{\theta}\vert h,\mathcal{M}\r)$, using the Metropolis-Hastings sampler \cite{Metropolis53Equation,Hastings70Monte,newman1999monte}. Specifically, we perform random jumps in the logarithm space of the parameters ($\log \kappa_{ij}$ and $\log \beta_k$). The jump kernel was chosen to be uniformly distributed $\text{Unif}\l(-D,D\r)$, where the meta-parameter   $D$ (diffusivity) globally regulates how wide the isotropic diffusion kernel is. For each model structure, we adjusted the meta-parameter $D$ such that the acceptance rate of the Metropolis-Hastings sampler was between $0.2$ and $0.3$ \cite{roberts1997Weak}. We assumed that our prior is uniform in the logarithm space $\log \theta_i$. 

Before running the MCMC samplers, we randomized $400$ initial guesses of the model parameters and forward evolved the MCMC for $5\times 10^4$ iterations each chain. Most of the chains converged to a unique parameter region and the likelihood value in this region was significantly higher than the few chains trapped in (presumably) local maxima. We independently initiated $32$ MCMC chains with different random initial seeds from the parameter values that maximized the likelihood in the previous test runs. We collected a total length (sum of the length of all $32$ chains, $>10^7$) to accurately approximate the posterior distribution $\mathbb{P} \l(\vec{\theta} \vert h, \mathcal{M}\r)$. \\

\noindent{\bf Computing the evidence $\mathbb{P}\l(h \vert \mathcal{M}\r)$.} As described in \cite{Robert2009Computational}, the evidence $\mathbb{P}\l(h \vert \mathcal{M}\r)$ is computed by the algebraic identity
\eq{
\mathbb{P}\l(h \vert \mathcal{M}\r) = \l[\int \frac{\phi\l(\vec{\theta'}\r)\mathbb{P}\l(\vec{\theta'}\vert h,\mathcal{M}\r)}{\mathbb{P}\l(h\vert \vec{\theta'}, M\r) \mathbb{P}\l(\vec{\theta'}\vert \mathcal{M} \r)}\dd \vec{\theta'} \r]^{-1},
}{}
with an importance sampler $\phi \l(\vec{\theta'}\r)$ satisfying the normalization condition
\eq{
\int \phi \l(\vec{\theta'}\r) \dd \vec{\theta'} = 1.
}{}
In this work, the posterior distributions were exclusively unimodal. Therefore, we chose the importance sampler to be proportional to an indicator function on an ellipsoid located at the high posterior density region. We first ranked the posterior chain by their likelihood and selected the top 20\% parameter sets to construct the importance sampler. We performed a Principle Component Analysis on the selected samples to compute the mean $\bar{\theta_k}$, variance $\sigma_{k}^2$  and eigenvector $\hat{e}_{k}$, $k=1,2,\ldots 6$, in the six-dimensional parameter space. We used the eigenvalues and eigenvectors to construct an ellipsoid centering at the mean and with the axis length proportional to the eigenvalues along with the eigenvectors: 
\eq{
\mathcal{E}:= \l\{\vec{\theta} \l\vert \sum_{k,j=1}^6 \frac{\l(R_{k,j} \theta_j - R_{k,j} \bar{\theta}_j \r)^2}{\alpha \sigma_k^2} <1\r.\r\}
}{}
where $R_{i,j}:=\l(\hat{e}_i\r)_j$ is the linear transformation onto the eigenbasis. We tuned the meta-parameter $\alpha$ such that there were precisely 20\% of the points of the posterior chains inside the ellipsoid $\mathcal{E}$. These samples were then used to compute the evidence.

One must also specify the prior distribution $\mathbb{P}\l(\vec{\theta}\vert \mathcal{M}\r)$ to compute the evidence. For simplicity, we imposed a uniform prior in the logarithm space of the parameters, bounded by $\l(10^{-16}, 10^4\r)$. $\mathbb{P}\l(\vec{\theta}\vert \mathcal{M}\r)$is $1/V$, where $V$ is the bounded volume in the parameter space. Let the posterior chains to be $\l\{\vec{\theta}_k\r\}_{k=1}^{N_P}$, where $N_P$ is the total number of samples in the posterior chains. Then, the marginalized likelihood is estimated computed by the Monte Carlo sampler:
\eq{
\widehat{\mathbb{P}} \l(h \vert \mathcal{M} \r)= \l[\sum_{k=1}^{N_P} \frac{1}{\hat{\mathbb{P}}\l(h \vert \vec{\theta}_k, \mathcal{M}\r)} \mathbf{1}_{\vec{\theta}_k\in \mathcal{E}}\r],
}{}
where $\mathbf{1}_{\vec{\theta}_k\in \mathcal{E}}$ is the characteristic function which is equal to $1$ if $\vec{\theta}_k$ is in the ellipsoid $\mathcal{E}$, and $0$ otherwise. In Figure 5, we presented a normalized probability among all the 64 linear 3-state models we considered: 
\eq{
\bar{\mathbb{P}}\l(h\vert \mathcal{M}_i\r):= \frac{\widehat{\mathbb{P}}\l(h\vert \mathcal{M}_i\r)}{\sum_{j=1}^{64} \widehat{\mathbb{P}}\l(h\vert \mathcal{M}_j\r)},
}{}
which is reported in Fig.~\ref{fig:5}.

The Schwarz Index is an asymptotic result of the Bayesian evidence $\mathbb{P}\l(h\vert \mathcal{M}\r)$ when the sample size is large \cite{schwarz1978Estimating}. Given a model $\mathcal{M}_i$, the Bayesian Information Criterion (BIC) is defined to be twice of the its Schwarz Index
\eq{
\text{BIC}\l(\mathcal{M}_i\r):= -2 \mathcal{L}_i^{\max} + m_i \log N,
}{}
where $\mathcal{L}_i^{\max}$ and $m_i$ are maximum likelihood and the number of free parameters of model $\mathcal{M}_i$ respectively, and $N$ is the sample size (number of data). Thus, to estimate the normalized probability $\bar{\mathbb{P}}$ using BIC, we use 
\eq{
\bar{\mathbb{P}}_\text{BIC}\l(h\vert \mathcal{M}_i\r):= \frac{\exp \l[-\frac{1}{2}\text{BIC}\l(\mathcal{M}_i\r)\r]}{\sum_{j=1}^{64} \exp \l[-\frac{1}{2}\text{BIC}\l(\mathcal{M}_j\r)\r]}.
}{}
We remark that the calcaulation of BIC only involves estimating the maximum likelihood $\mathcal{L}_i^{\max}$ of each model $\mathcal{M}_i$ and not the full posterior distribution $\mathbb{P}\l(\vec{\theta}\vert \mathcal{M}_i\r)$.

Akaike Information Criterion (AIC) is another commonly adopted matrix information criterion. The motivation of AIC is to minimize the information loss, measured by the Kullback--Leibler divergence (KL) of the prediction from the data. It is derived \cite{Akaike1998Information} as
\eq{
\text{AIC}\l(\mathcal{M}_i\r):= -2 \mathcal{L}_i^{\max} + 2 m_i + 2 \frac{ m_i + m_i^2}{N- m_i - 1}.
}{}
Thus, the (normalized) evidence calculated by the AIC is
\eq{
\bar{\mathbb{P}}_\text{AIC}\l(h\vert \mathcal{M}_i\r):= \frac{\exp \l[-\frac{1}{2}\text{AIC}\l(\mathcal{M}_i\r)\r]}{\sum_{j=1}^{64} \exp \l[-\frac{1}{2}\text{AIC}\l(\mathcal{M}_j\r)\r]}.
}{}
We remark that the negative logarithm of the likelihood function \eqref{eq:likelihoodDef} converges to  $ N \times \text{KL}\l( h\l(\omega\r)/N \vert\vert \mathbb{P}\l(\omega\vert \mathcal{M}_i, \vec{\theta}\r)\r)$ only when $N\gg 1$ such that the Multinomial coefficient $\mathbb{M}_\ell$ can be expanded by the Stirling approximation. In most biological cases, the sample size is far from this regime, and the Kullback--Leibler divergence is a poor choice to approximate the likelihood function \eqref{eq:likelihoodDef}.


\begin{thebibliography}{61}%
\makeatletter
\providecommand \@ifxundefined [1]{%
 \@ifx{#1\undefined}
}%
\providecommand \@ifnum [1]{%
 \ifnum #1\expandafter \@firstoftwo
 \else \expandafter \@secondoftwo
 \fi
}%
\providecommand \@ifx [1]{%
 \ifx #1\expandafter \@firstoftwo
 \else \expandafter \@secondoftwo
 \fi
}%
\providecommand \natexlab [1]{#1}%
\providecommand \enquote  [1]{``#1''}%
\providecommand \bibnamefont  [1]{#1}%
\providecommand \bibfnamefont [1]{#1}%
\providecommand \citenamefont [1]{#1}%
\providecommand \href@noop [0]{\@secondoftwo}%
\providecommand \href [0]{\begingroup \@sanitize@url \@href}%
\providecommand \@href[1]{\@@startlink{#1}\@@href}%
\providecommand \@@href[1]{\endgroup#1\@@endlink}%
\providecommand \@sanitize@url [0]{\catcode `\\12\catcode `\$12\catcode
  `\&12\catcode `\#12\catcode `\^12\catcode `\_12\catcode `\%12\relax}%
\providecommand \@@startlink[1]{}%
\providecommand \@@endlink[0]{}%
\providecommand \url  [0]{\begingroup\@sanitize@url \@url }%
\providecommand \@url [1]{\endgroup\@href {#1}{\urlprefix }}%
\providecommand \urlprefix  [0]{URL }%
\providecommand \Eprint [0]{\href }%
\providecommand \doibase [0]{http://dx.doi.org/}%
\providecommand \selectlanguage [0]{\@gobble}%
\providecommand \bibinfo  [0]{\@secondoftwo}%
\providecommand \bibfield  [0]{\@secondoftwo}%
\providecommand \translation [1]{[#1]}%
\providecommand \BibitemOpen [0]{}%
\providecommand \bibitemStop [0]{}%
\providecommand \bibitemNoStop [0]{.\EOS\space}%
\providecommand \EOS [0]{\spacefactor3000\relax}%
\providecommand \BibitemShut  [1]{\csname bibitem#1\endcsname}%
\let\auto@bib@innerbib\@empty
\bibitem [{\citenamefont {Pichon}\ \emph {et~al.}(2018)\citenamefont {Pichon},
  \citenamefont {Lagha}, \citenamefont {Mueller},\ and\ \citenamefont
  {Bertrand}}]{pichon2018}%
  \BibitemOpen
  \bibfield  {author} {\bibinfo {author} {\bibfnamefont {X.}~\bibnamefont
  {Pichon}}, \bibinfo {author} {\bibfnamefont {M.}~\bibnamefont {Lagha}},
  \bibinfo {author} {\bibfnamefont {F.}~\bibnamefont {Mueller}}, \ and\
  \bibinfo {author} {\bibfnamefont {E.}~\bibnamefont {Bertrand}},\ }\href@noop
  {} {\bibfield  {journal} {\bibinfo  {journal} {Mol. Cell}\ }\textbf {\bibinfo
  {volume} {71}},\ \bibinfo {pages} {468} (\bibinfo {year} {2018})}\BibitemShut
  {NoStop}%
\bibitem [{\citenamefont {Golding}\ \emph {et~al.}(2005)\citenamefont
  {Golding}, \citenamefont {Paulsson}, \citenamefont {Zawilski},\ and\
  \citenamefont {Cox}}]{golding2005}%
  \BibitemOpen
  \bibfield  {author} {\bibinfo {author} {\bibfnamefont {I.}~\bibnamefont
  {Golding}}, \bibinfo {author} {\bibfnamefont {J.}~\bibnamefont {Paulsson}},
  \bibinfo {author} {\bibfnamefont {S.~M.}\ \bibnamefont {Zawilski}}, \ and\
  \bibinfo {author} {\bibfnamefont {E.~C.}\ \bibnamefont {Cox}},\ }\href@noop
  {} {\bibfield  {journal} {\bibinfo  {journal} {Cell}\ }\textbf {\bibinfo
  {volume} {123}},\ \bibinfo {pages} {1025} (\bibinfo {year}
  {2005})}\BibitemShut {NoStop}%
\bibitem [{\citenamefont {Larson}\ \emph {et~al.}(2011)\citenamefont {Larson},
  \citenamefont {Zenklusen}, \citenamefont {Wu}, \citenamefont {Chao},\ and\
  \citenamefont {Singer}}]{larson2011}%
  \BibitemOpen
  \bibfield  {author} {\bibinfo {author} {\bibfnamefont {D.~R.}\ \bibnamefont
  {Larson}}, \bibinfo {author} {\bibfnamefont {D.}~\bibnamefont {Zenklusen}},
  \bibinfo {author} {\bibfnamefont {B.}~\bibnamefont {Wu}}, \bibinfo {author}
  {\bibfnamefont {J.~A.}\ \bibnamefont {Chao}}, \ and\ \bibinfo {author}
  {\bibfnamefont {R.~H.}\ \bibnamefont {Singer}},\ }\href@noop {} {\bibfield
  {journal} {\bibinfo  {journal} {Science}\ }\textbf {\bibinfo {volume}
  {332}},\ \bibinfo {pages} {475} (\bibinfo {year} {2011})}\BibitemShut
  {NoStop}%
\bibitem [{\citenamefont {Corrigan}\ \emph {et~al.}(2016)\citenamefont
  {Corrigan}, \citenamefont {Tunnacliffe}, \citenamefont {Cannon},\ and\
  \citenamefont {Chubb}}]{corrigan2016}%
  \BibitemOpen
  \bibfield  {author} {\bibinfo {author} {\bibfnamefont {A.~M.}\ \bibnamefont
  {Corrigan}}, \bibinfo {author} {\bibfnamefont {E.}~\bibnamefont
  {Tunnacliffe}}, \bibinfo {author} {\bibfnamefont {D.}~\bibnamefont {Cannon}},
  \ and\ \bibinfo {author} {\bibfnamefont {J.~R.}\ \bibnamefont {Chubb}},\
  }\href@noop {} {\bibfield  {journal} {\bibinfo  {journal} {eLife}\ }\textbf
  {\bibinfo {volume} {5}} (\bibinfo {year} {2016})}\BibitemShut {NoStop}%
\bibitem [{\citenamefont {Fukaya}\ \emph {et~al.}(2016)\citenamefont {Fukaya},
  \citenamefont {Lim},\ and\ \citenamefont {Levine}}]{fukaya2016}%
  \BibitemOpen
  \bibfield  {author} {\bibinfo {author} {\bibfnamefont {T.}~\bibnamefont
  {Fukaya}}, \bibinfo {author} {\bibfnamefont {B.}~\bibnamefont {Lim}}, \ and\
  \bibinfo {author} {\bibfnamefont {M.}~\bibnamefont {Levine}},\ }\href@noop {}
  {\bibfield  {journal} {\bibinfo  {journal} {Cell}\ }\textbf {\bibinfo
  {volume} {166}},\ \bibinfo {pages} {358} (\bibinfo {year}
  {2016})}\BibitemShut {NoStop}%
\bibitem [{\citenamefont {Ozbudak}\ \emph {et~al.}(2002)\citenamefont
  {Ozbudak}, \citenamefont {Thattai}, \citenamefont {Kurtser}, \citenamefont
  {Grossman},\ and\ \citenamefont {van Oudenaarden}}]{ozbudak2002}%
  \BibitemOpen
  \bibfield  {author} {\bibinfo {author} {\bibfnamefont {E.~M.}\ \bibnamefont
  {Ozbudak}}, \bibinfo {author} {\bibfnamefont {M.}~\bibnamefont {Thattai}},
  \bibinfo {author} {\bibfnamefont {I.}~\bibnamefont {Kurtser}}, \bibinfo
  {author} {\bibfnamefont {A.~D.}\ \bibnamefont {Grossman}}, \ and\ \bibinfo
  {author} {\bibfnamefont {A.}~\bibnamefont {van Oudenaarden}},\ }\href@noop {}
  {\bibfield  {journal} {\bibinfo  {journal} {Nat. Genet.}\ }\textbf {\bibinfo
  {volume} {31}},\ \bibinfo {pages} {69} (\bibinfo {year} {2002})}\BibitemShut
  {NoStop}%
\bibitem [{\citenamefont {Elowitz}\ \emph {et~al.}(2002)\citenamefont
  {Elowitz}, \citenamefont {Levine}, \citenamefont {Siggia},\ and\
  \citenamefont {Swain}}]{elowitz2002}%
  \BibitemOpen
  \bibfield  {author} {\bibinfo {author} {\bibfnamefont {M.~B.}\ \bibnamefont
  {Elowitz}}, \bibinfo {author} {\bibfnamefont {A.~J.}\ \bibnamefont {Levine}},
  \bibinfo {author} {\bibfnamefont {E.~D.}\ \bibnamefont {Siggia}}, \ and\
  \bibinfo {author} {\bibfnamefont {P.~S.}\ \bibnamefont {Swain}},\ }\href@noop
  {} {\bibfield  {journal} {\bibinfo  {journal} {Science}\ }\textbf {\bibinfo
  {volume} {297}},\ \bibinfo {pages} {1183} (\bibinfo {year}
  {2002})}\BibitemShut {NoStop}%
\bibitem [{\citenamefont {Suter}\ \emph {et~al.}(2011)\citenamefont {Suter},
  \citenamefont {Molina}, \citenamefont {Gatfield}, \citenamefont {Schneider},
  \citenamefont {Schibler},\ and\ \citenamefont {Naef}}]{suter2011}%
  \BibitemOpen
  \bibfield  {author} {\bibinfo {author} {\bibfnamefont {D.~M.}\ \bibnamefont
  {Suter}}, \bibinfo {author} {\bibfnamefont {N.}~\bibnamefont {Molina}},
  \bibinfo {author} {\bibfnamefont {D.}~\bibnamefont {Gatfield}}, \bibinfo
  {author} {\bibfnamefont {K.}~\bibnamefont {Schneider}}, \bibinfo {author}
  {\bibfnamefont {U.}~\bibnamefont {Schibler}}, \ and\ \bibinfo {author}
  {\bibfnamefont {F.}~\bibnamefont {Naef}},\ }\href@noop {} {\bibfield
  {journal} {\bibinfo  {journal} {Science}\ }\textbf {\bibinfo {volume}
  {332}},\ \bibinfo {pages} {472} (\bibinfo {year} {2011})}\BibitemShut
  {NoStop}%
\bibitem [{\citenamefont {Nicolas}\ \emph {et~al.}(2018)\citenamefont
  {Nicolas}, \citenamefont {Zoller}, \citenamefont {Suter},\ and\ \citenamefont
  {Naef}}]{nicolas2018}%
  \BibitemOpen
  \bibfield  {author} {\bibinfo {author} {\bibfnamefont {D.}~\bibnamefont
  {Nicolas}}, \bibinfo {author} {\bibfnamefont {B.}~\bibnamefont {Zoller}},
  \bibinfo {author} {\bibfnamefont {D.~M.}\ \bibnamefont {Suter}}, \ and\
  \bibinfo {author} {\bibfnamefont {F.}~\bibnamefont {Naef}},\ }\href@noop {}
  {\bibfield  {journal} {\bibinfo  {journal} {Proc. Natl. Acad. Sci. U.S.A.}\
  }\textbf {\bibinfo {volume} {115}},\ \bibinfo {pages} {7153} (\bibinfo {year}
  {2018})}\BibitemShut {NoStop}%
\bibitem [{\citenamefont {Raj}\ \emph {et~al.}(2008)\citenamefont {Raj},
  \citenamefont {van~den Bogaard}, \citenamefont {Rifkin}, \citenamefont {van
  Oudenaarden},\ and\ \citenamefont {Tyagi}}]{raj2008}%
  \BibitemOpen
  \bibfield  {author} {\bibinfo {author} {\bibfnamefont {A.}~\bibnamefont
  {Raj}}, \bibinfo {author} {\bibfnamefont {P.}~\bibnamefont {van~den
  Bogaard}}, \bibinfo {author} {\bibfnamefont {S.~A.}\ \bibnamefont {Rifkin}},
  \bibinfo {author} {\bibfnamefont {A.}~\bibnamefont {van Oudenaarden}}, \ and\
  \bibinfo {author} {\bibfnamefont {S.}~\bibnamefont {Tyagi}},\ }\href@noop {}
  {\bibfield  {journal} {\bibinfo  {journal} {Nat. Methods}\ }\textbf {\bibinfo
  {volume} {5}},\ \bibinfo {pages} {877} (\bibinfo {year} {2008})}\BibitemShut
  {NoStop}%
\bibitem [{\citenamefont {Zenklusen}\ \emph {et~al.}(2008)\citenamefont
  {Zenklusen}, \citenamefont {Larson},\ and\ \citenamefont
  {Singer}}]{zenklusen2008}%
  \BibitemOpen
  \bibfield  {author} {\bibinfo {author} {\bibfnamefont {D.}~\bibnamefont
  {Zenklusen}}, \bibinfo {author} {\bibfnamefont {D.~R.}\ \bibnamefont
  {Larson}}, \ and\ \bibinfo {author} {\bibfnamefont {R.~H.}\ \bibnamefont
  {Singer}},\ }\href@noop {} {\bibfield  {journal} {\bibinfo  {journal} {Nat.
  Struct. Mol. Biol.}\ }\textbf {\bibinfo {volume} {15}},\ \bibinfo {pages}
  {1263} (\bibinfo {year} {2008})}\BibitemShut {NoStop}%
\bibitem [{\citenamefont {Tang}\ \emph {et~al.}(2009)\citenamefont {Tang},
  \citenamefont {Barbacioru}, \citenamefont {Wang}, \citenamefont {Nordman},
  \citenamefont {Lee}, \citenamefont {Xu}, \citenamefont {Wang}, \citenamefont
  {Bodeau}, \citenamefont {Tuch}, \citenamefont {Siddiqui}, \citenamefont
  {Lao},\ and\ \citenamefont {Surani}}]{tang2009}%
  \BibitemOpen
  \bibfield  {author} {\bibinfo {author} {\bibfnamefont {F.}~\bibnamefont
  {Tang}}, \bibinfo {author} {\bibfnamefont {C.}~\bibnamefont {Barbacioru}},
  \bibinfo {author} {\bibfnamefont {Y.}~\bibnamefont {Wang}}, \bibinfo {author}
  {\bibfnamefont {E.}~\bibnamefont {Nordman}}, \bibinfo {author} {\bibfnamefont
  {C.}~\bibnamefont {Lee}}, \bibinfo {author} {\bibfnamefont {N.}~\bibnamefont
  {Xu}}, \bibinfo {author} {\bibfnamefont {X.}~\bibnamefont {Wang}}, \bibinfo
  {author} {\bibfnamefont {J.}~\bibnamefont {Bodeau}}, \bibinfo {author}
  {\bibfnamefont {B.~B.}\ \bibnamefont {Tuch}}, \bibinfo {author}
  {\bibfnamefont {A.}~\bibnamefont {Siddiqui}}, \bibinfo {author}
  {\bibfnamefont {K.}~\bibnamefont {Lao}}, \ and\ \bibinfo {author}
  {\bibfnamefont {M.~A.}\ \bibnamefont {Surani}},\ }\href@noop {} {\bibfield
  {journal} {\bibinfo  {journal} {Nat. Methods}\ }\textbf {\bibinfo {volume}
  {6}},\ \bibinfo {pages} {377} (\bibinfo {year} {2009})}\BibitemShut {NoStop}%
\bibitem [{\citenamefont {Hwang}\ \emph {et~al.}(2018)\citenamefont {Hwang},
  \citenamefont {Lee},\ and\ \citenamefont {Bang}}]{hwang2018}%
  \BibitemOpen
  \bibfield  {author} {\bibinfo {author} {\bibfnamefont {B.}~\bibnamefont
  {Hwang}}, \bibinfo {author} {\bibfnamefont {J.~H.}\ \bibnamefont {Lee}}, \
  and\ \bibinfo {author} {\bibfnamefont {D.}~\bibnamefont {Bang}},\ }\href@noop
  {} {\bibfield  {journal} {\bibinfo  {journal} {Exp. Mol. Med.}\ }\textbf
  {\bibinfo {volume} {50}},\ \bibinfo {pages} {96} (\bibinfo {year}
  {2018})}\BibitemShut {NoStop}%
\bibitem [{\citenamefont {Munsky}\ \emph {et~al.}(2015)\citenamefont {Munsky},
  \citenamefont {Fox},\ and\ \citenamefont {Neuert}}]{munsky2015}%
  \BibitemOpen
  \bibfield  {author} {\bibinfo {author} {\bibfnamefont {B.}~\bibnamefont
  {Munsky}}, \bibinfo {author} {\bibfnamefont {Z.}~\bibnamefont {Fox}}, \ and\
  \bibinfo {author} {\bibfnamefont {G.}~\bibnamefont {Neuert}},\ }\href@noop {}
  {\bibfield  {journal} {\bibinfo  {journal} {Methods}\ }\textbf {\bibinfo
  {volume} {85}},\ \bibinfo {pages} {12} (\bibinfo {year} {2015})}\BibitemShut
  {NoStop}%
\bibitem [{\citenamefont {G{\'o}mez-Schiavon}\ \emph
  {et~al.}(2017)\citenamefont {G{\'o}mez-Schiavon}, \citenamefont {Chen},
  \citenamefont {West},\ and\ \citenamefont {Buchler}}]{Gomez-Schiavon2017}%
  \BibitemOpen
  \bibfield  {author} {\bibinfo {author} {\bibfnamefont {M.}~\bibnamefont
  {G{\'o}mez-Schiavon}}, \bibinfo {author} {\bibfnamefont {L.-F.}\ \bibnamefont
  {Chen}}, \bibinfo {author} {\bibfnamefont {A.~E.}\ \bibnamefont {West}}, \
  and\ \bibinfo {author} {\bibfnamefont {N.~E.}\ \bibnamefont {Buchler}},\
  }\href@noop {} {\bibfield  {journal} {\bibinfo  {journal} {Genome Biology}\
  }\textbf {\bibinfo {volume} {18}},\ \bibinfo {pages} {164} (\bibinfo {year}
  {2017})}\BibitemShut {NoStop}%
\bibitem [{\citenamefont {Raj}\ and\ \citenamefont {van
  Oudenaarden}(2008)}]{rajcell2008}%
  \BibitemOpen
  \bibfield  {author} {\bibinfo {author} {\bibfnamefont {A.}~\bibnamefont
  {Raj}}\ and\ \bibinfo {author} {\bibfnamefont {A.}~\bibnamefont {van
  Oudenaarden}},\ }\href@noop {} {\bibfield  {journal} {\bibinfo  {journal}
  {Cell}\ }\textbf {\bibinfo {volume} {135}},\ \bibinfo {pages} {216} (\bibinfo
  {year} {2008})}\BibitemShut {NoStop}%
\bibitem [{\citenamefont {Nicolas}\ \emph {et~al.}(2017)\citenamefont
  {Nicolas}, \citenamefont {Phillips},\ and\ \citenamefont
  {Naef}}]{nicolas2017}%
  \BibitemOpen
  \bibfield  {author} {\bibinfo {author} {\bibfnamefont {D.}~\bibnamefont
  {Nicolas}}, \bibinfo {author} {\bibfnamefont {N.~E.}\ \bibnamefont
  {Phillips}}, \ and\ \bibinfo {author} {\bibfnamefont {F.}~\bibnamefont
  {Naef}},\ }\href@noop {} {\bibfield  {journal} {\bibinfo  {journal} {Mol
  Biosyst}\ }\textbf {\bibinfo {volume} {13}},\ \bibinfo {pages} {1280}
  (\bibinfo {year} {2017})}\BibitemShut {NoStop}%
\bibitem [{\citenamefont {Shahrezaei}\ and\ \citenamefont
  {Swain}(2008)}]{Swain2008Analytical}%
  \BibitemOpen
  \bibfield  {author} {\bibinfo {author} {\bibfnamefont {V.}~\bibnamefont
  {Shahrezaei}}\ and\ \bibinfo {author} {\bibfnamefont {P.~S.}\ \bibnamefont
  {Swain}},\ }\href {\doibase 10.1073/pnas.0803850105} {\bibfield  {journal}
  {\bibinfo  {journal} {Proceedings of the National Academy of Sciences}\
  }\textbf {\bibinfo {volume} {105}},\ \bibinfo {pages} {17256} (\bibinfo
  {year} {2008})}\BibitemShut {NoStop}%
\bibitem [{\citenamefont {Kumar}\ \emph {et~al.}(2014)\citenamefont {Kumar},
  \citenamefont {Platini},\ and\ \citenamefont {Kulkarni}}]{Kumar2014Exact}%
  \BibitemOpen
  \bibfield  {author} {\bibinfo {author} {\bibfnamefont {N.}~\bibnamefont
  {Kumar}}, \bibinfo {author} {\bibfnamefont {T.}~\bibnamefont {Platini}}, \
  and\ \bibinfo {author} {\bibfnamefont {R.~V.}\ \bibnamefont {Kulkarni}},\
  }\href@noop {} {\bibfield  {journal} {\bibinfo  {journal} {Phys. Rev. Lett.}\
  }\textbf {\bibinfo {volume} {113}},\ \bibinfo {pages} {268105} (\bibinfo
  {year} {2014})}\BibitemShut {NoStop}%
\bibitem [{\citenamefont {Tiberi}\ \emph {et~al.}(2018)\citenamefont {Tiberi},
  \citenamefont {Walsh}, \citenamefont {Cavallaro}, \citenamefont
  {Hebenstreit},\ and\ \citenamefont {Finkenstädt}}]{Tiberi2018Bayesian}%
  \BibitemOpen
  \bibfield  {author} {\bibinfo {author} {\bibfnamefont {S.}~\bibnamefont
  {Tiberi}}, \bibinfo {author} {\bibfnamefont {M.}~\bibnamefont {Walsh}},
  \bibinfo {author} {\bibfnamefont {M.}~\bibnamefont {Cavallaro}}, \bibinfo
  {author} {\bibfnamefont {D.}~\bibnamefont {Hebenstreit}}, \ and\ \bibinfo
  {author} {\bibfnamefont {B.}~\bibnamefont {Finkenstädt}},\ }\href@noop {}
  {\bibfield  {journal} {\bibinfo  {journal} {Bioinformatics}\ }\textbf
  {\bibinfo {volume} {34}},\ \bibinfo {pages} {i647} (\bibinfo {year}
  {2018})}\BibitemShut {NoStop}%
\bibitem [{\citenamefont {Munsky}\ and\ \citenamefont
  {Khammash}(2006)}]{Munsky2006FSP}%
  \BibitemOpen
  \bibfield  {author} {\bibinfo {author} {\bibfnamefont {B.}~\bibnamefont
  {Munsky}}\ and\ \bibinfo {author} {\bibfnamefont {M.}~\bibnamefont
  {Khammash}},\ }\href {\doibase 10.1063/1.2145882} {\bibfield  {journal}
  {\bibinfo  {journal} {The Journal of Chemical Physics}\ }\textbf {\bibinfo
  {volume} {124}},\ \bibinfo {pages} {044104} (\bibinfo {year}
  {2006})}\BibitemShut {NoStop}%
\bibitem [{\citenamefont {Cao}\ \emph {et~al.}(2016)\citenamefont {Cao},
  \citenamefont {Terebus},\ and\ \citenamefont {Liang}}]{cao2016ACME}%
  \BibitemOpen
  \bibfield  {author} {\bibinfo {author} {\bibfnamefont {Y.}~\bibnamefont
  {Cao}}, \bibinfo {author} {\bibfnamefont {A.}~\bibnamefont {Terebus}}, \ and\
  \bibinfo {author} {\bibfnamefont {J.}~\bibnamefont {Liang}},\ }\href
  {\doibase 10.1137/15M1034180} {\bibfield  {journal} {\bibinfo  {journal}
  {Multiscale Modeling \& Simulation}\ }\textbf {\bibinfo {volume} {14}},\
  \bibinfo {pages} {923} (\bibinfo {year} {2016})}\BibitemShut {NoStop}%
\bibitem [{\citenamefont {Raj}\ \emph {et~al.}(2006)\citenamefont {Raj},
  \citenamefont {Peskin}, \citenamefont {Tranchina}, \citenamefont {Vargas},\
  and\ \citenamefont {Tyagi}}]{raj2006}%
  \BibitemOpen
  \bibfield  {author} {\bibinfo {author} {\bibfnamefont {A.}~\bibnamefont
  {Raj}}, \bibinfo {author} {\bibfnamefont {C.~S.}\ \bibnamefont {Peskin}},
  \bibinfo {author} {\bibfnamefont {D.}~\bibnamefont {Tranchina}}, \bibinfo
  {author} {\bibfnamefont {D.~Y.}\ \bibnamefont {Vargas}}, \ and\ \bibinfo
  {author} {\bibfnamefont {S.}~\bibnamefont {Tyagi}},\ }\href@noop {}
  {\bibfield  {journal} {\bibinfo  {journal} {PLoS Biol.}\ }\textbf {\bibinfo
  {volume} {4}},\ \bibinfo {pages} {e309} (\bibinfo {year} {2006})}\BibitemShut
  {NoStop}%
\bibitem [{\citenamefont {Little}\ \emph {et~al.}(2013)\citenamefont {Little},
  \citenamefont {Tikhonov},\ and\ \citenamefont {Gregor}}]{little2013}%
  \BibitemOpen
  \bibfield  {author} {\bibinfo {author} {\bibfnamefont {S.~C.}\ \bibnamefont
  {Little}}, \bibinfo {author} {\bibfnamefont {M.}~\bibnamefont {Tikhonov}}, \
  and\ \bibinfo {author} {\bibfnamefont {T.}~\bibnamefont {Gregor}},\
  }\href@noop {} {\bibfield  {journal} {\bibinfo  {journal} {Cell}\ }\textbf
  {\bibinfo {volume} {154}},\ \bibinfo {pages} {789} (\bibinfo {year}
  {2013})}\BibitemShut {NoStop}%
\bibitem [{\citenamefont {Bahar~Halpern}\ \emph {et~al.}(2015)\citenamefont
  {Bahar~Halpern}, \citenamefont {Tanami}, \citenamefont {Landen},
  \citenamefont {Chapal}, \citenamefont {Szlak}, \citenamefont {Hutzler},
  \citenamefont {Nizhberg},\ and\ \citenamefont
  {Itzkovitz}}]{bahar-halpern2015}%
  \BibitemOpen
  \bibfield  {author} {\bibinfo {author} {\bibfnamefont {K.}~\bibnamefont
  {Bahar~Halpern}}, \bibinfo {author} {\bibfnamefont {S.}~\bibnamefont
  {Tanami}}, \bibinfo {author} {\bibfnamefont {S.}~\bibnamefont {Landen}},
  \bibinfo {author} {\bibfnamefont {M.}~\bibnamefont {Chapal}}, \bibinfo
  {author} {\bibfnamefont {L.}~\bibnamefont {Szlak}}, \bibinfo {author}
  {\bibfnamefont {A.}~\bibnamefont {Hutzler}}, \bibinfo {author} {\bibfnamefont
  {A.}~\bibnamefont {Nizhberg}}, \ and\ \bibinfo {author} {\bibfnamefont
  {S.}~\bibnamefont {Itzkovitz}},\ }\href@noop {} {\bibfield  {journal}
  {\bibinfo  {journal} {Mol. Cell}\ }\textbf {\bibinfo {volume} {58}},\
  \bibinfo {pages} {147} (\bibinfo {year} {2015})}\BibitemShut {NoStop}%
\bibitem [{\citenamefont {Skinner}\ \emph {et~al.}(2016)\citenamefont
  {Skinner}, \citenamefont {Xu}, \citenamefont {Nagarkar-Jaiswal},
  \citenamefont {Freire}, \citenamefont {Zwaka},\ and\ \citenamefont
  {Golding}}]{skinner2016}%
  \BibitemOpen
  \bibfield  {author} {\bibinfo {author} {\bibfnamefont {S.~O.}\ \bibnamefont
  {Skinner}}, \bibinfo {author} {\bibfnamefont {H.}~\bibnamefont {Xu}},
  \bibinfo {author} {\bibfnamefont {S.}~\bibnamefont {Nagarkar-Jaiswal}},
  \bibinfo {author} {\bibfnamefont {P.~R.}\ \bibnamefont {Freire}}, \bibinfo
  {author} {\bibfnamefont {T.~P.}\ \bibnamefont {Zwaka}}, \ and\ \bibinfo
  {author} {\bibfnamefont {I.}~\bibnamefont {Golding}},\ }\href@noop {}
  {\bibfield  {journal} {\bibinfo  {journal} {eLife}\ }\textbf {\bibinfo
  {volume} {5}},\ \bibinfo {pages} {e12175} (\bibinfo {year}
  {2016})}\BibitemShut {NoStop}%
\bibitem [{\citenamefont {Gillespie}(1976)}]{Gillespie76General}%
  \BibitemOpen
  \bibfield  {author} {\bibinfo {author} {\bibfnamefont {D.~T.}\ \bibnamefont
  {Gillespie}},\ }\href@noop {} {\bibfield  {journal} {\bibinfo  {journal}
  {Journal of Computational Physics}\ }\textbf {\bibinfo {volume} {22}},\
  \bibinfo {pages} {403 } (\bibinfo {year} {1976})}\BibitemShut {NoStop}%
\bibitem [{\citenamefont {Gillespie}(1977)}]{Gillespie77Exact}%
  \BibitemOpen
  \bibfield  {author} {\bibinfo {author} {\bibfnamefont {D.~T.}\ \bibnamefont
  {Gillespie}},\ }\href {\doibase 10.1021/j100540a008} {\bibfield  {journal}
  {\bibinfo  {journal} {The Journal of Physical Chemistry}\ }\textbf {\bibinfo
  {volume} {81}},\ \bibinfo {pages} {2340} (\bibinfo {year}
  {1977})}\BibitemShut {NoStop}%
\bibitem [{\citenamefont {MacKay}(2005)}]{MacKay2005Information}%
  \BibitemOpen
  \bibfield  {author} {\bibinfo {author} {\bibfnamefont {D.~J.~C.}\
  \bibnamefont {MacKay}},\ }\href@noop {} {\emph {\bibinfo {title} {Information
  Theory, Inference, and Learning Algorithms}}}\ (\bibinfo  {publisher}
  {Cambridge University Press},\ \bibinfo {year} {2005})\BibitemShut {NoStop}%
\bibitem [{\citenamefont {Sivia}\ and\ \citenamefont
  {Skilling}(2006)}]{sivia2006data}%
  \BibitemOpen
  \bibfield  {author} {\bibinfo {author} {\bibfnamefont {D.}~\bibnamefont
  {Sivia}}\ and\ \bibinfo {author} {\bibfnamefont {J.}~\bibnamefont
  {Skilling}},\ }\href@noop {} {\emph {\bibinfo {title} {Data analysis: a
  Bayesian tutorial}}}\ (\bibinfo  {publisher} {OUP Oxford},\ \bibinfo {year}
  {2006})\BibitemShut {NoStop}%
\bibitem [{\citenamefont {Green}(1995)}]{Green1995Reversible}%
  \BibitemOpen
  \bibfield  {author} {\bibinfo {author} {\bibfnamefont {P.~J.}\ \bibnamefont
  {Green}},\ }\href@noop {} {\bibfield  {journal} {\bibinfo  {journal}
  {Biometrika}\ }\textbf {\bibinfo {volume} {82}},\ \bibinfo {pages} {711}
  (\bibinfo {year} {1995})}\BibitemShut {NoStop}%
\bibitem [{\citenamefont {Kass}\ and\ \citenamefont
  {Raftery}(1995)}]{Kass1995BayesFactors}%
  \BibitemOpen
  \bibfield  {author} {\bibinfo {author} {\bibfnamefont {R.~E.}\ \bibnamefont
  {Kass}}\ and\ \bibinfo {author} {\bibfnamefont {A.~E.}\ \bibnamefont
  {Raftery}},\ }\href {\doibase 10.1080/01621459.1995.10476572} {\bibfield
  {journal} {\bibinfo  {journal} {Journal of the American Statistical
  Association}\ }\textbf {\bibinfo {volume} {90}},\ \bibinfo {pages} {773}
  (\bibinfo {year} {1995})}\BibitemShut {NoStop}%
\bibitem [{\citenamefont {Weinberg}(2012)}]{Weinberg2012Computing}%
  \BibitemOpen
  \bibfield  {author} {\bibinfo {author} {\bibfnamefont {M.~D.}\ \bibnamefont
  {Weinberg}},\ }\href {\doibase 10.1214/12-BA725} {\bibfield  {journal}
  {\bibinfo  {journal} {Bayesian Anal.}\ }\textbf {\bibinfo {volume} {7}},\
  \bibinfo {pages} {737} (\bibinfo {year} {2012})}\BibitemShut {NoStop}%
\bibitem [{\citenamefont {Pajor}(2017)}]{Pajor2017Estimating}%
  \BibitemOpen
  \bibfield  {author} {\bibinfo {author} {\bibfnamefont {A.}~\bibnamefont
  {Pajor}},\ }\href {\doibase 10.1214/16-BA1001} {\bibfield  {journal}
  {\bibinfo  {journal} {Bayesian Anal.}\ }\textbf {\bibinfo {volume} {12}},\
  \bibinfo {pages} {261} (\bibinfo {year} {2017})}\BibitemShut {NoStop}%
\bibitem [{\citenamefont {Robert}\ and\ \citenamefont
  {Wraith}(2009)}]{Robert2009Computational}%
  \BibitemOpen
  \bibfield  {author} {\bibinfo {author} {\bibfnamefont {C.~P.}\ \bibnamefont
  {Robert}}\ and\ \bibinfo {author} {\bibfnamefont {D.}~\bibnamefont
  {Wraith}},\ }\href {\doibase 10.1063/1.3275622} {\bibfield  {journal}
  {\bibinfo  {journal} {AIP Conference Proceedings}\ }\textbf {\bibinfo
  {volume} {1193}},\ \bibinfo {pages} {251} (\bibinfo {year}
  {2009})}\BibitemShut {NoStop}%
\bibitem [{\citenamefont {Bokes}\ \emph {et~al.}(2013)\citenamefont {Bokes},
  \citenamefont {King}, \citenamefont {Wood},\ and\ \citenamefont
  {Loose}}]{Bokes2013Transcriptional}%
  \BibitemOpen
  \bibfield  {author} {\bibinfo {author} {\bibfnamefont {P.}~\bibnamefont
  {Bokes}}, \bibinfo {author} {\bibfnamefont {J.~R.}\ \bibnamefont {King}},
  \bibinfo {author} {\bibfnamefont {A.~T.~A.}\ \bibnamefont {Wood}}, \ and\
  \bibinfo {author} {\bibfnamefont {M.}~\bibnamefont {Loose}},\ }\href
  {\doibase 10.1007/s11538-013-9811-z} {\bibfield  {journal} {\bibinfo
  {journal} {Bulletin of Mathematical Biology}\ }\textbf {\bibinfo {volume}
  {75}},\ \bibinfo {pages} {351} (\bibinfo {year} {2013})}\BibitemShut
  {NoStop}%
\bibitem [{\citenamefont {Lin}\ and\ \citenamefont
  {Doering}(2016)}]{lin2016gene}%
  \BibitemOpen
  \bibfield  {author} {\bibinfo {author} {\bibfnamefont {Y.~T.}\ \bibnamefont
  {Lin}}\ and\ \bibinfo {author} {\bibfnamefont {C.~R.}\ \bibnamefont
  {Doering}},\ }\href {\doibase 10.1103/PhysRevE.93.022409} {\bibfield
  {journal} {\bibinfo  {journal} {Phys. Rev. E}\ }\textbf {\bibinfo {volume}
  {93}},\ \bibinfo {pages} {022409} (\bibinfo {year} {2016})}\BibitemShut
  {NoStop}%
\bibitem [{\citenamefont {Lin}\ and\ \citenamefont
  {Galla}(2016)}]{lin2016bursting}%
  \BibitemOpen
  \bibfield  {author} {\bibinfo {author} {\bibfnamefont {Y.~T.}\ \bibnamefont
  {Lin}}\ and\ \bibinfo {author} {\bibfnamefont {T.}~\bibnamefont {Galla}},\
  }\href {\doibase 10.1098/rsif.2015.0772} {\bibfield  {journal} {\bibinfo
  {journal} {Journal of The Royal Society Interface}\ }\textbf {\bibinfo
  {volume} {13}} (\bibinfo {year} {2016}),\ 10.1098/rsif.2015.0772}\BibitemShut
  {NoStop}%
\bibitem [{\citenamefont {Hufton}\ \emph {et~al.}(2016)\citenamefont {Hufton},
  \citenamefont {Lin}, \citenamefont {Galla},\ and\ \citenamefont
  {Mckane}}]{Hufton2017}%
  \BibitemOpen
  \bibfield  {author} {\bibinfo {author} {\bibfnamefont {P.}~\bibnamefont
  {Hufton}}, \bibinfo {author} {\bibfnamefont {Y.}~\bibnamefont {Lin}},
  \bibinfo {author} {\bibfnamefont {T.}~\bibnamefont {Galla}}, \ and\ \bibinfo
  {author} {\bibfnamefont {A.}~\bibnamefont {Mckane}},\ }\href {\doibase
  10.1103/PhysRevE.93.052119} {\bibfield  {journal} {\bibinfo  {journal}
  {Physical Review E: covering statistical, nonlinear, biological, and soft
  matter physics}\ }\textbf {\bibinfo {volume} {93}} (\bibinfo {year} {2016}),\
  10.1103/PhysRevE.93.052119}\BibitemShut {NoStop}%
\bibitem [{\citenamefont {Bressloff}(2017)}]{Bressloff2017stochastic}%
  \BibitemOpen
  \bibfield  {author} {\bibinfo {author} {\bibfnamefont {P.~C.}\ \bibnamefont
  {Bressloff}},\ }\href@noop {} {\bibfield  {journal} {\bibinfo  {journal}
  {Journal of Physics A: Mathematical and Theoretical}\ }\textbf {\bibinfo
  {volume} {50}},\ \bibinfo {pages} {133001} (\bibinfo {year}
  {2017})}\BibitemShut {NoStop}%
\bibitem [{\citenamefont {Lin}\ \emph {et~al.}(2018)\citenamefont {Lin},
  \citenamefont {Hufton}, \citenamefont {Lee},\ and\ \citenamefont
  {Potoyan}}]{Lin2018Stochastic}%
  \BibitemOpen
  \bibfield  {author} {\bibinfo {author} {\bibfnamefont {Y.~T.}\ \bibnamefont
  {Lin}}, \bibinfo {author} {\bibfnamefont {P.~G.}\ \bibnamefont {Hufton}},
  \bibinfo {author} {\bibfnamefont {E.~J.}\ \bibnamefont {Lee}}, \ and\
  \bibinfo {author} {\bibfnamefont {D.~A.}\ \bibnamefont {Potoyan}},\ }\href
  {\doibase 10.1371/journal.pcbi.1006000} {\bibfield  {journal} {\bibinfo
  {journal} {PLOS Computational Biology}\ }\textbf {\bibinfo {volume} {14}},\
  \bibinfo {pages} {1} (\bibinfo {year} {2018})}\BibitemShut {NoStop}%
\bibitem [{\citenamefont {Lin}\ and\ \citenamefont
  {Buchler}(2018)}]{lin2018Efficient}%
  \BibitemOpen
  \bibfield  {author} {\bibinfo {author} {\bibfnamefont {Y.~T.}\ \bibnamefont
  {Lin}}\ and\ \bibinfo {author} {\bibfnamefont {N.~E.}\ \bibnamefont
  {Buchler}},\ }\href {\doibase 10.1098/rsif.2017.0804} {\bibfield  {journal}
  {\bibinfo  {journal} {Journal of The Royal Society Interface}\ }\textbf
  {\bibinfo {volume} {15}} (\bibinfo {year} {2018}),\
  10.1098/rsif.2017.0804}\BibitemShut {NoStop}%
\bibitem [{\citenamefont {{Herbach}}(2018)}]{Herbach2018}%
  \BibitemOpen
  \bibfield  {author} {\bibinfo {author} {\bibfnamefont {U.}~\bibnamefont
  {{Herbach}}},\ }\href@noop {} {\bibfield  {journal} {\bibinfo  {journal}
  {ArXiv e-prints}\ } (\bibinfo {year} {2018})},\ \Eprint
  {http://arxiv.org/abs/1804.05740} {arXiv:1804.05740 [math.PR]} \BibitemShut
  {NoStop}%
\bibitem [{\citenamefont {Wang}\ \emph {et~al.}(2018)\citenamefont {Wang},
  \citenamefont {Huang}, \citenamefont {Torre}, \citenamefont {Dueck},
  \citenamefont {Shaffer}, \citenamefont {Murray}, \citenamefont {Raj},
  \citenamefont {Li},\ and\ \citenamefont {Zhang}}]{Wang2018Gene}%
  \BibitemOpen
  \bibfield  {author} {\bibinfo {author} {\bibfnamefont {J.}~\bibnamefont
  {Wang}}, \bibinfo {author} {\bibfnamefont {M.}~\bibnamefont {Huang}},
  \bibinfo {author} {\bibfnamefont {E.}~\bibnamefont {Torre}}, \bibinfo
  {author} {\bibfnamefont {H.}~\bibnamefont {Dueck}}, \bibinfo {author}
  {\bibfnamefont {S.}~\bibnamefont {Shaffer}}, \bibinfo {author} {\bibfnamefont
  {J.}~\bibnamefont {Murray}}, \bibinfo {author} {\bibfnamefont
  {A.}~\bibnamefont {Raj}}, \bibinfo {author} {\bibfnamefont {M.}~\bibnamefont
  {Li}}, \ and\ \bibinfo {author} {\bibfnamefont {N.~R.}\ \bibnamefont
  {Zhang}},\ }\href {\doibase 10.1073/pnas.1721085115} {\bibfield  {journal}
  {\bibinfo  {journal} {Proceedings of the National Academy of Sciences}\
  }\textbf {\bibinfo {volume} {115}},\ \bibinfo {pages} {E6437} (\bibinfo
  {year} {2018})}\BibitemShut {NoStop}%
\bibitem [{\citenamefont {Munsky}\ \emph {et~al.}(2018)\citenamefont {Munsky},
  \citenamefont {Li}, \citenamefont {Fox}, \citenamefont {Shepherd},\ and\
  \citenamefont {Neuert}}]{Munsky2018Distribution}%
  \BibitemOpen
  \bibfield  {author} {\bibinfo {author} {\bibfnamefont {B.}~\bibnamefont
  {Munsky}}, \bibinfo {author} {\bibfnamefont {G.}~\bibnamefont {Li}}, \bibinfo
  {author} {\bibfnamefont {Z.~R.}\ \bibnamefont {Fox}}, \bibinfo {author}
  {\bibfnamefont {D.~P.}\ \bibnamefont {Shepherd}}, \ and\ \bibinfo {author}
  {\bibfnamefont {G.}~\bibnamefont {Neuert}},\ }\href {\doibase
  10.1073/pnas.1804060115} {\bibfield  {journal} {\bibinfo  {journal}
  {Proceedings of the National Academy of Sciences}\ }\textbf {\bibinfo
  {volume} {115}},\ \bibinfo {pages} {7533} (\bibinfo {year}
  {2018})}\BibitemShut {NoStop}%
\bibitem [{\citenamefont {Duane}\ \emph {et~al.}(1987)\citenamefont {Duane},
  \citenamefont {Kennedy}, \citenamefont {Pendleton},\ and\ \citenamefont
  {Roweth}}]{DUANE1987216}%
  \BibitemOpen
  \bibfield  {author} {\bibinfo {author} {\bibfnamefont {S.}~\bibnamefont
  {Duane}}, \bibinfo {author} {\bibfnamefont {A.}~\bibnamefont {Kennedy}},
  \bibinfo {author} {\bibfnamefont {B.~J.}\ \bibnamefont {Pendleton}}, \ and\
  \bibinfo {author} {\bibfnamefont {D.}~\bibnamefont {Roweth}},\ }\href
  {\doibase https://doi.org/10.1016/0370-2693(87)91197-X} {\bibfield  {journal}
  {\bibinfo  {journal} {Physics Letters B}\ }\textbf {\bibinfo {volume}
  {195}},\ \bibinfo {pages} {216 } (\bibinfo {year} {1987})}\BibitemShut
  {NoStop}%
\bibitem [{\citenamefont {Neal}\ \emph {et~al.}(2011)\citenamefont {Neal} \emph
  {et~al.}}]{neal2011mcmc}%
  \BibitemOpen
  \bibfield  {author} {\bibinfo {author} {\bibfnamefont {R.~M.}\ \bibnamefont
  {Neal}} \emph {et~al.},\ }\href@noop {} {\bibfield  {journal} {\bibinfo
  {journal} {Handbook of Markov Chain Monte Carlo}\ }\textbf {\bibinfo {volume}
  {2}},\ \bibinfo {pages} {2} (\bibinfo {year} {2011})}\BibitemShut {NoStop}%
\bibitem [{\citenamefont {Betancourt}(2017)}]{betancourt2017conceptual}%
  \BibitemOpen
  \bibfield  {author} {\bibinfo {author} {\bibfnamefont {M.}~\bibnamefont
  {Betancourt}},\ }\href@noop {} {\bibfield  {journal} {\bibinfo  {journal}
  {arXiv preprint arXiv:1701.02434}\ } (\bibinfo {year} {2017})}\BibitemShut
  {NoStop}%
\bibitem [{\citenamefont {Toni}\ \emph {et~al.}(2009)\citenamefont {Toni},
  \citenamefont {Welch}, \citenamefont {Strelkowa}, \citenamefont {Ipsen},\
  and\ \citenamefont {Stumpf}}]{Toni187Approximate}%
  \BibitemOpen
  \bibfield  {author} {\bibinfo {author} {\bibfnamefont {T.}~\bibnamefont
  {Toni}}, \bibinfo {author} {\bibfnamefont {D.}~\bibnamefont {Welch}},
  \bibinfo {author} {\bibfnamefont {N.}~\bibnamefont {Strelkowa}}, \bibinfo
  {author} {\bibfnamefont {A.}~\bibnamefont {Ipsen}}, \ and\ \bibinfo {author}
  {\bibfnamefont {M.~P.}\ \bibnamefont {Stumpf}},\ }\href {\doibase
  10.1098/rsif.2008.0172} {\bibfield  {journal} {\bibinfo  {journal} {Journal
  of The Royal Society Interface}\ }\textbf {\bibinfo {volume} {6}},\ \bibinfo
  {pages} {187} (\bibinfo {year} {2009})}\BibitemShut {NoStop}%
\bibitem [{\citenamefont {Kampen}(2007)}]{vankampen2007}%
  \BibitemOpen
  \bibfield  {author} {\bibinfo {author} {\bibfnamefont {N.~V.}\ \bibnamefont
  {Kampen}},\ }\href@noop {} {\emph {\bibinfo {title} {Stochastic processes in
  physics and chemistry}}}\ (\bibinfo  {publisher} {North Holland},\ \bibinfo
  {year} {2007})\BibitemShut {NoStop}%
\bibitem [{\citenamefont {Davis}(1984)}]{Davis1983Piecewise}%
  \BibitemOpen
  \bibfield  {author} {\bibinfo {author} {\bibfnamefont {M.~H.~A.}\
  \bibnamefont {Davis}},\ }\href@noop {} {\bibfield  {journal} {\bibinfo
  {journal} {Journal of the Royal Statistical Society. Series B
  (Methodological)}\ }\textbf {\bibinfo {volume} {46}},\ \bibinfo {pages} {353}
  (\bibinfo {year} {1984})}\BibitemShut {NoStop}%
\bibitem [{\citenamefont {Bena}(2006)}]{Bena2006Dichotomous}%
  \BibitemOpen
  \bibfield  {author} {\bibinfo {author} {\bibfnamefont {I.}~\bibnamefont
  {Bena}},\ }\href {\doibase 10.1142/S0217979206034881} {\bibfield  {journal}
  {\bibinfo  {journal} {International Journal of Modern Physics B}\ }\textbf
  {\bibinfo {volume} {20}},\ \bibinfo {pages} {2825} (\bibinfo {year}
  {2006})}\BibitemShut {NoStop}%
\bibitem [{\citenamefont {Faggionato}\ \emph {et~al.}(2009)\citenamefont
  {Faggionato}, \citenamefont {Gabrielli},\ and\ \citenamefont
  {Ribezzi~Crivellari}}]{Faggionato2009NonEquilibrium}%
  \BibitemOpen
  \bibfield  {author} {\bibinfo {author} {\bibfnamefont {A.}~\bibnamefont
  {Faggionato}}, \bibinfo {author} {\bibfnamefont {D.}~\bibnamefont
  {Gabrielli}}, \ and\ \bibinfo {author} {\bibfnamefont {M.}~\bibnamefont
  {Ribezzi~Crivellari}},\ }\href {\doibase 10.1007/s10955-009-9850-x}
  {\bibfield  {journal} {\bibinfo  {journal} {Journal of Statistical Physics}\
  }\textbf {\bibinfo {volume} {137}},\ \bibinfo {pages} {259} (\bibinfo {year}
  {2009})}\BibitemShut {NoStop}%
\bibitem [{\citenamefont {Neuert}\ \emph {et~al.}(2013)\citenamefont {Neuert},
  \citenamefont {Munsky}, \citenamefont {Tan}, \citenamefont {Teytelman},
  \citenamefont {Khammash},\ and\ \citenamefont {van
  Oudenaarden}}]{neuert2013}%
  \BibitemOpen
  \bibfield  {author} {\bibinfo {author} {\bibfnamefont {G.}~\bibnamefont
  {Neuert}}, \bibinfo {author} {\bibfnamefont {B.}~\bibnamefont {Munsky}},
  \bibinfo {author} {\bibfnamefont {R.~Z.}\ \bibnamefont {Tan}}, \bibinfo
  {author} {\bibfnamefont {L.}~\bibnamefont {Teytelman}}, \bibinfo {author}
  {\bibfnamefont {M.}~\bibnamefont {Khammash}}, \ and\ \bibinfo {author}
  {\bibfnamefont {A.}~\bibnamefont {van Oudenaarden}},\ }\href@noop {}
  {\bibfield  {journal} {\bibinfo  {journal} {Science}\ }\textbf {\bibinfo
  {volume} {339}},\ \bibinfo {pages} {584} (\bibinfo {year}
  {2013})}\BibitemShut {NoStop}%
\bibitem [{\citenamefont {Senecal}\ \emph {et~al.}(2014)\citenamefont
  {Senecal}, \citenamefont {Munsky}, \citenamefont {Proux}, \citenamefont {Ly},
  \citenamefont {Braye}, \citenamefont {Zimmer}, \citenamefont {Mueller},\ and\
  \citenamefont {Darzacq}}]{senecal2014}%
  \BibitemOpen
  \bibfield  {author} {\bibinfo {author} {\bibfnamefont {A.}~\bibnamefont
  {Senecal}}, \bibinfo {author} {\bibfnamefont {B.}~\bibnamefont {Munsky}},
  \bibinfo {author} {\bibfnamefont {F.}~\bibnamefont {Proux}}, \bibinfo
  {author} {\bibfnamefont {N.}~\bibnamefont {Ly}}, \bibinfo {author}
  {\bibfnamefont {F.~E.}\ \bibnamefont {Braye}}, \bibinfo {author}
  {\bibfnamefont {C.}~\bibnamefont {Zimmer}}, \bibinfo {author} {\bibfnamefont
  {F.}~\bibnamefont {Mueller}}, \ and\ \bibinfo {author} {\bibfnamefont
  {X.}~\bibnamefont {Darzacq}},\ }\href@noop {} {\bibfield  {journal} {\bibinfo
   {journal} {Cell Rep}\ }\textbf {\bibinfo {volume} {8}},\ \bibinfo {pages}
  {75} (\bibinfo {year} {2014})}\BibitemShut {NoStop}%
\bibitem [{\citenamefont {Metropolis}\ \emph {et~al.}(1953)\citenamefont
  {Metropolis}, \citenamefont {Rosenbluth}, \citenamefont {Rosenbluth},
  \citenamefont {Teller},\ and\ \citenamefont {Teller}}]{Metropolis53Equation}%
  \BibitemOpen
  \bibfield  {author} {\bibinfo {author} {\bibfnamefont {N.}~\bibnamefont
  {Metropolis}}, \bibinfo {author} {\bibfnamefont {A.~W.}\ \bibnamefont
  {Rosenbluth}}, \bibinfo {author} {\bibfnamefont {M.~N.}\ \bibnamefont
  {Rosenbluth}}, \bibinfo {author} {\bibfnamefont {A.~H.}\ \bibnamefont
  {Teller}}, \ and\ \bibinfo {author} {\bibfnamefont {E.}~\bibnamefont
  {Teller}},\ }\href {\doibase 10.1063/1.1699114} {\bibfield  {journal}
  {\bibinfo  {journal} {The Journal of Chemical Physics}\ }\textbf {\bibinfo
  {volume} {21}},\ \bibinfo {pages} {1087} (\bibinfo {year}
  {1953})}\BibitemShut {NoStop}%
\bibitem [{\citenamefont {Hastings}(1970)}]{Hastings70Monte}%
  \BibitemOpen
  \bibfield  {author} {\bibinfo {author} {\bibfnamefont {W.~K.}\ \bibnamefont
  {Hastings}},\ }\href {\doibase 10.1093/biomet/57.1.97} {\bibfield  {journal}
  {\bibinfo  {journal} {Biometrika}\ }\textbf {\bibinfo {volume} {57}},\
  \bibinfo {pages} {97} (\bibinfo {year} {1970})}\BibitemShut {NoStop}%
\bibitem [{\citenamefont {Newman}\ and\ \citenamefont
  {Barkema}(1999)}]{newman1999monte}%
  \BibitemOpen
  \bibfield  {author} {\bibinfo {author} {\bibfnamefont {M.}~\bibnamefont
  {Newman}}\ and\ \bibinfo {author} {\bibfnamefont {G.}~\bibnamefont
  {Barkema}},\ }\href@noop {} {\emph {\bibinfo {title} {Monte carlo methods in
  statistical physics chapter 1-4}}}\ (\bibinfo  {publisher} {Oxford University
  Press: New York, USA},\ \bibinfo {year} {1999})\BibitemShut {NoStop}%
\bibitem [{\citenamefont {Roberts}\ \emph {et~al.}(1997)\citenamefont
  {Roberts}, \citenamefont {Gelman},\ and\ \citenamefont
  {Gilks}}]{roberts1997Weak}%
  \BibitemOpen
  \bibfield  {author} {\bibinfo {author} {\bibfnamefont {G.~O.}\ \bibnamefont
  {Roberts}}, \bibinfo {author} {\bibfnamefont {A.}~\bibnamefont {Gelman}}, \
  and\ \bibinfo {author} {\bibfnamefont {W.~R.}\ \bibnamefont {Gilks}},\ }\href
  {\doibase 10.1214/aoap/1034625254} {\bibfield  {journal} {\bibinfo  {journal}
  {Ann. Appl. Probab.}\ }\textbf {\bibinfo {volume} {7}},\ \bibinfo {pages}
  {110} (\bibinfo {year} {1997})}\BibitemShut {NoStop}%
\bibitem [{\citenamefont {Schwarz}(1978)}]{schwarz1978Estimating}%
  \BibitemOpen
  \bibfield  {author} {\bibinfo {author} {\bibfnamefont {G.}~\bibnamefont
  {Schwarz}},\ }\href {\doibase 10.1214/aos/1176344136} {\bibfield  {journal}
  {\bibinfo  {journal} {Ann. Statist.}\ }\textbf {\bibinfo {volume} {6}},\
  \bibinfo {pages} {461} (\bibinfo {year} {1978})}\BibitemShut {NoStop}%
\bibitem [{\citenamefont {Akaike}(1998)}]{Akaike1998Information}%
  \BibitemOpen
  \bibfield  {author} {\bibinfo {author} {\bibfnamefont {H.}~\bibnamefont
  {Akaike}},\ }\enquote {\bibinfo {title} {Information theory and an extension
  of the maximum likelihood principle},}\ in\ \href {\doibase
  10.1007/978-1-4612-1694-0_15} {\emph {\bibinfo {booktitle} {Selected Papers
  of Hirotugu Akaike}}},\ \bibinfo {editor} {edited by\ \bibinfo {editor}
  {\bibfnamefont {E.}~\bibnamefont {Parzen}}, \bibinfo {editor} {\bibfnamefont
  {K.}~\bibnamefont {Tanabe}}, \ and\ \bibinfo {editor} {\bibfnamefont
  {G.}~\bibnamefont {Kitagawa}}}\ (\bibinfo  {publisher} {Springer New York},\
  \bibinfo {address} {New York, NY},\ \bibinfo {year} {1998})\ pp.\ \bibinfo
  {pages} {199--213}\BibitemShut {NoStop}%
\end{thebibliography}
\end{document}